\documentclass[pra,twocolumn,superscriptaddress,amsmath,amssymb,longbibliography]{revtex4-1}
\usepackage{hyperref}
\pdfoutput=1
\usepackage{graphicx}
\usepackage{multirow}
\usepackage{array}
\usepackage{amsmath}
\usepackage{bm}
\usepackage{amssymb}
\usepackage{textcomp}
\usepackage{stmaryrd}
\usepackage{ulem}
\usepackage{gensymb}

\usepackage{import}
\usepackage{color}
\usepackage{verbatim}

\usepackage{float}

\usepackage{sistyle} 

\def\XXint#1#2#3{{\setbox0=\hbox{$#1{#2#3}{\int}$}
     \vcenter{\hbox{$#2#3$}}\kern-.5\wd0}}

\makeatletter
\DeclareRobustCommand{\element}[1]{\@element#1\@nil}
\def\@element#1#2\@nil{%
  #1%
  \if\relax#2\relax\else\MakeLowercase{#2}\fi}
\makeatother

\begin{document}

\title{Filtering the photoluminescence spectra of atomically thin semiconductors with graphene}

\author{Etienne Lorchat}
\altaffiliation{These authors contributed equally to this study}
\affiliation{Universit\'e de Strasbourg, CNRS, Institut de Physique et Chimie des Mat\'eriaux de Strasbourg (IPCMS), UMR 7504, F-67000 Strasbourg, France}

\author{Luis E. Parra L\'{o}pez}
\altaffiliation{These authors contributed equally to this study}
\affiliation{Universit\'e de Strasbourg, CNRS, Institut de Physique et Chimie des Mat\'eriaux de Strasbourg (IPCMS), UMR 7504, F-67000 Strasbourg, France}

\author{C\'edric Robert}
\affiliation{Universit\'e de Toulouse, INSA-CNRS-UPS, LPCNO, 135 Avenue de Rangueil, 31077, Toulouse, France}

\author{Delphine Lagarde}
\affiliation{Universit\'e de Toulouse, INSA-CNRS-UPS, LPCNO, 135 Avenue de Rangueil, 31077, Toulouse, France}

\author{Guillaume Froehlicher}
\affiliation{Universit\'e de Strasbourg, CNRS, Institut de Physique et Chimie des Mat\'eriaux de Strasbourg (IPCMS), UMR 7504, F-67000 Strasbourg, France}

\author{Takashi Taniguchi}
\affiliation{National Institute for Materials Science, Tsukuba, Ibaraki 305-0044, Japan}

\author{Kenji Watanabe}
\affiliation{National Institute for Materials Science, Tsukuba, Ibaraki 305-0044, Japan}

\author{Xavier Marie}
\affiliation{Universit\'e de Toulouse, INSA-CNRS-UPS, LPCNO, 135 Avenue de Rangueil, 31077, Toulouse, France}
\affiliation{Institut Universitaire de France, 1 rue Descartes, 75231 Paris cedex 05, France}

\author{St\'ephane Berciaud}
\email{stephane.berciaud@ipcms.unistra.fr}
\affiliation{Universit\'e de Strasbourg, CNRS, Institut de Physique et Chimie des Mat\'eriaux de Strasbourg (IPCMS), UMR 7504, F-67000 Strasbourg, France}
\affiliation{Institut Universitaire de France, 1 rue Descartes, 75231 Paris cedex 05, France}

\begin{abstract}
Atomically thin semiconductors made from transition metal dichalcogenides (TMDs) are model systems for investigations of strong light-matter interactions and applications in nanophotonics, opto-electronics and valley-tronics. However, the photoluminescence spectra of TMD monolayers display a large number of features that are particularly challenging to decipher. On a practical level, monochromatic TMD-based emitters would be beneficial for low-dimensional devices but this challenge is yet to be resolved. Here, we show that graphene, directly stacked onto TMD monolayers enables single and narrow-line photoluminescence arising solely from TMD neutral excitons. This filtering effect stems from complete neutralization of the TMD by graphene combined with selective non-radiative transfer of long-lived excitonic species to graphene. Our approach is applied to four tungsten and molybdenum-based TMDs and establishes TMD/graphene heterostructures as a unique set of opto-electronic building blocks, suitable for electroluminescent systems emitting visible and near-infrared photons at near THz rate with linewidths approaching the lifetime limit.

\end{abstract}

\maketitle

 TMD monolayers (thereafter simply denoted TMD), such as MoS$_2$, MoSe$_2$, WS$_2$, WSe$_2$ are direct-bandgap semiconductors~\cite{Mak2010,Splendiani2010},  featuring short Bohr radii, large exciton binding energy of hundreds of meV~\cite{Wang2018,Goryca2019} and picosecond excitonic radiative lifetimes at low temperature~\cite{Palummo2015,Robert2016,Fang2019}, all arising from their strong 2D Coulomb interactions, reduced dielectric screening and large effective masses~\cite{Wang2018,Goryca2019}.  Since the first investigations of light emission from TMDs, it has been clear that their low-temperature spectra is composed of at least two  prominent features, stemming from bright neutral excitons ($\rm X^0$) and charged excitons (trions, $\rm X^{\star}$) \cite{Mak2013,Ross2013,Courtade2017} endowed with a binding energy of typically 20 to 40~meV relative to $\rm X^0$. Among the vast family of TMDs, one may distinguish between so-called dark and bright materials~\cite{Echeverry2016}. In the case of Tungsten-based TMDs, a spin-dark state lies lower than $\rm X^0$. Conversely, in Molybdenum based-TMDs, $\rm X^0$ lies below (MoSe$_2$) or very near (MoS$_2$) the spin-dark state, resulting in brighter emission at low temperature. As a result, $\rm X^0$ and $\rm X^{\star}$ emission dominate the PL spectrum of Mo-based TMDs~\cite{Ross2013}, whereas the emission spectra of W-based TMDs display a complex series of lines stemming from $\rm X^0$, bi-excitons ($\rm{XX}^0$)\cite{Barbone2018,Ye2018,Paur2019,Li2018}, charged excitonic states (including $\rm X^{\star}$~\cite{Courtade2017,Vaclavkova2018} and charged biexcitons ($\rm{XX}^{\star}$)~\cite{Barbone2018,Ye2018,Paur2019,Li2018}), spin-dark excitons~\cite{Robert2017,Zhou2017,Zhang2017}, defect-induced emission and exciton-phonon sidebands~\cite{Lindlau2018,Liu2019}.

Considerable progress has been made to deterministically observe intrinsic TMD emission features. In particular, encapsulation of TMDs in hexagonal boron nitride (BN) films results in narrower neutral exciton linewidth~\cite{Cadiz2017,Ajayi2017}, approaching the radiative limit~\cite{Fang2019,Scuri2018,Back2018}, without however, getting rid of the other emission features mentioned above. Even in electrostatically gated devices tuned near the charge neutrality point, sizeable emission sidebands remain observable at energies close to the $\rm X^{\star}$ feature, suggesting residual charge inhomogeneity~\cite{Courtade2017,Ross2013} or intrinsic contributions from longer-lived exciton-phonon replicas~\cite{Lindlau2018,Liu2019}.

\begin{figure*}[!th]
\begin{center}
\includegraphics[width=0.7\linewidth]{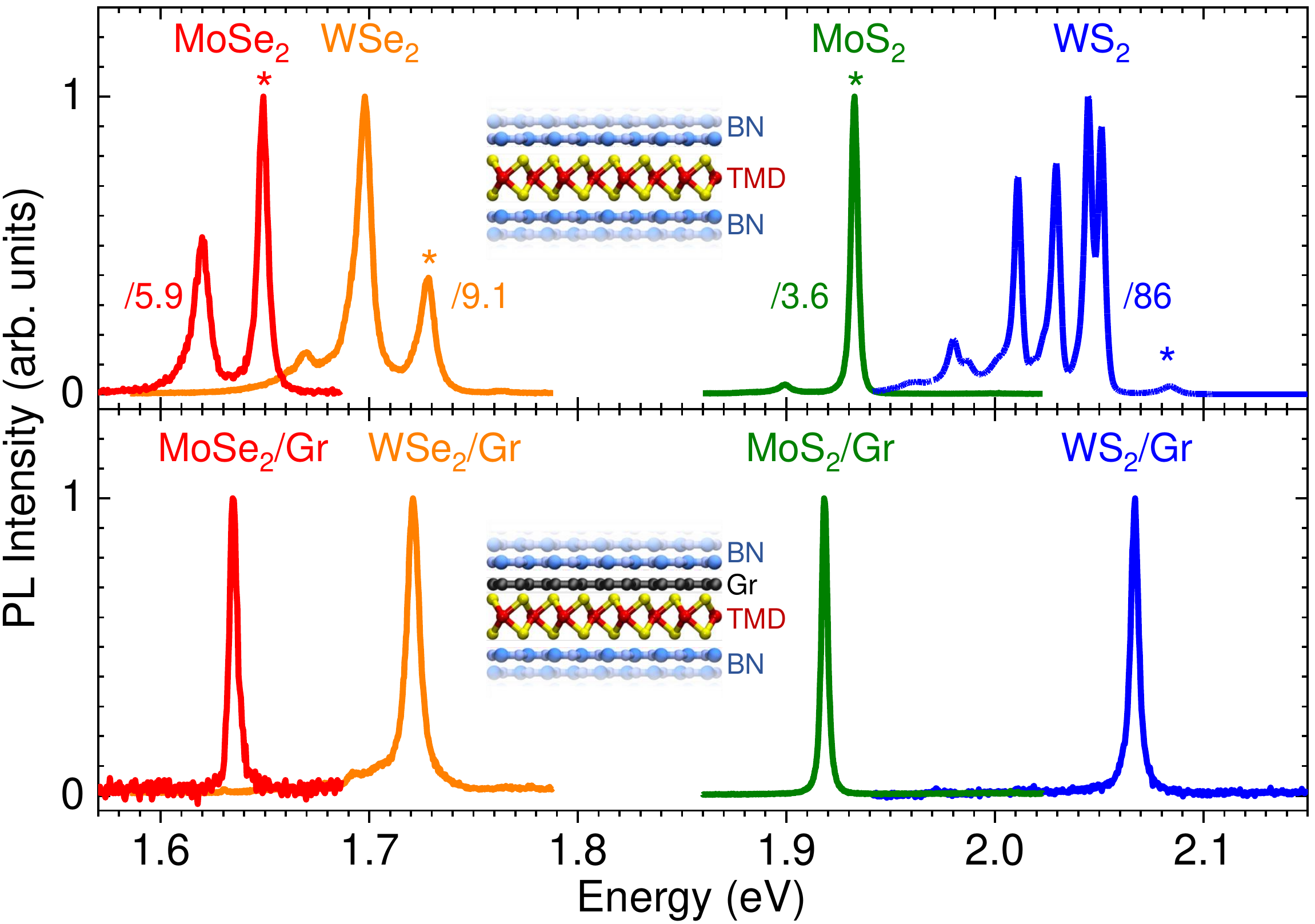}
\caption{\textbf{Graphene as a versatile emission filter.} Bright, single, and narrow-line photoluminescence (PL) spectra in TMD/graphene heterostructures are shown in the lower panel.  The PL spectra of neighboring BN-capped TMD regions are shown for comparison in the upper panel. A sketch of the van der Waals heterostructure is shown in each panel.  All spectra are normalized to unity relative to their most intense feature. The scaling factors indicated for each material allow direct comparison between the PL intensities in BN-capped TMD and BN-capped TMD/graphene. The bright neutral exciton ($\rm X^0$) lines in bare TMD are indicated by asterisks. All measurements were performed at temperatures below 20~K, in the linear regime  with continuous wave laser excitation at 2.33~eV (MoSe$_2$, MoS$_2$, WS$_2$) or 1.96~eV (WSe$_2$).}
\label{Fig1}
\end{center}
\end{figure*}

The complex emission spectra of TMD stimulate lively scientific debates. Conversely, obtaining atomically-thin semiconductors with single, narrow emission lines remains an important challenge in the field. An appealing solution could consist in interfacing a TMD monolayer with graphene. Indeed, the semi-metallic character of graphene and its highly symmetric electronic structure~\cite{Castroneto2009}, with its Dirac point lying within the bandgap of Mo- and W- based TMD~\cite{Wilson2017}, makes it an ideal electron and hole acceptor, through static charge transfer~\cite{Hill2017,Froehlicher2018}. Unfortunately, at room temperature, the \textit{effective} $\rm X^0$ lifetime is in the ns range~\cite{Palummo2015,Robert2016,Froehlicher2018} and interlayer coupling between TMDs and graphene results in significant PL quenching~\cite{He2014,Yuan2018,Froehlicher2018} due to picosecond energy transfer mediated by either charge tunneling (Dexter-type) or longer-range dipole-dipole interaction (F\"orster-type)~\cite{Froehlicher2018,Selig2019,Basko1999}. However, a much more favorable situation may occur at lower temperatures (typically, below 100~K), where the radiative lifetime of $\rm X^0$ drastically shortens~\cite{Robert2016,Palummo2015} and becomes of the same order of magnitude as the theoretically estimated energy transfer time~\cite{Selig2019}.

In this work, we demonstrate that W- and Mo-based TMDs coupled to a graphene monolayer exhibit only one single and narrow emission line that is assigned to $\rm X^0$ radiative recombination, indicating complete neutrality. The short-lived $\rm X^0$ states are minimally affected by non-radiative transfer to graphene and subsequent PL quenching, in stark contrast with longer-lived excitonic species, which are massively quenched. Graphene has been recognised as a partner material of choice to improve the opto-electronic response of TMDs~\cite{Massicotte2016}, whereas TMDs hold promise to improve  spin transport in graphene~\cite{Luo2017,Avsar2017}. Our results now establish TMD/graphene heterostructures as an outstanding light-emitting system readily interfaced with a quasi-transparent conductive channel.


\begin{figure*}[!ht]
\begin{center}
\includegraphics[width=0.8\linewidth]{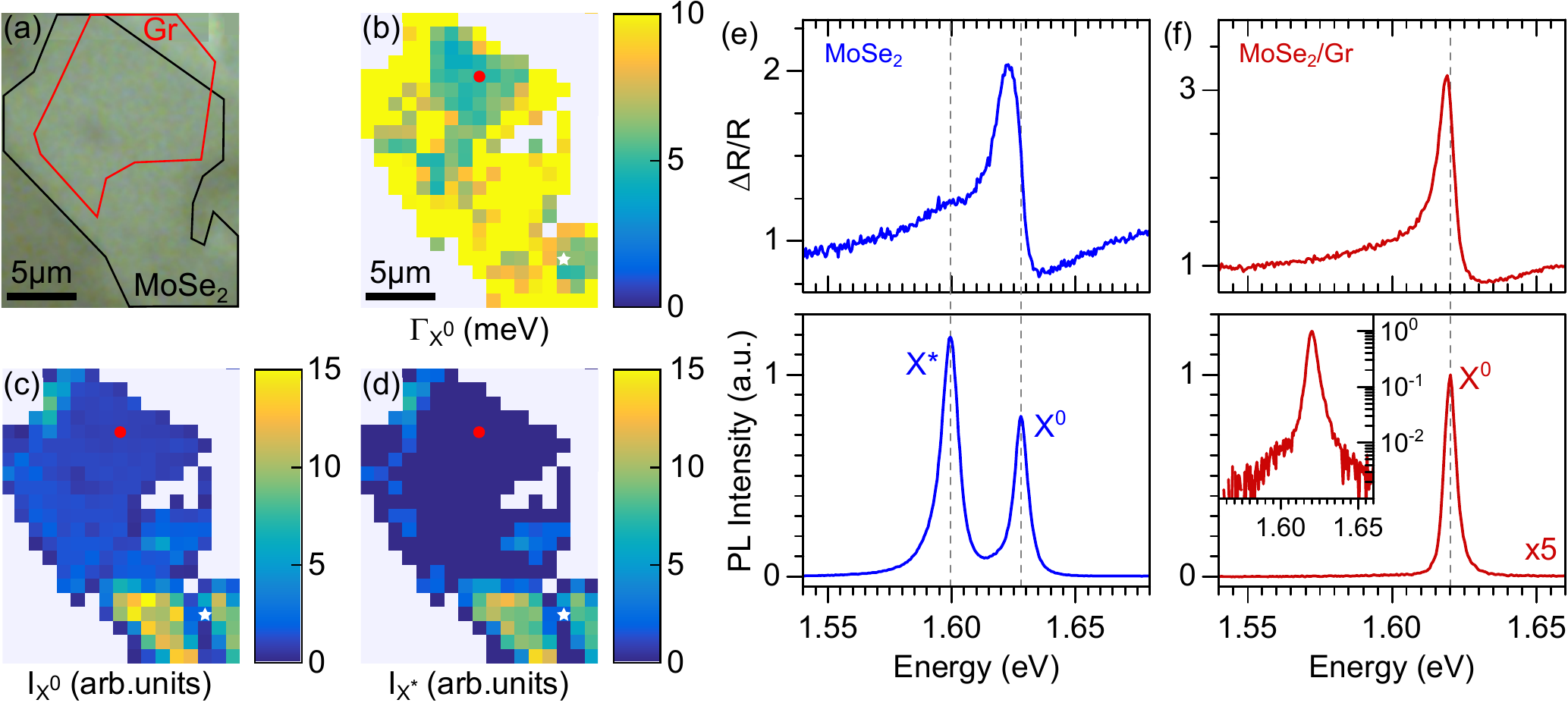}
\caption{\textbf{Neutralizing an atomically thin semiconductor with graphene.} (a) Optical image of a BN-capped MoSe$_2$/graphene sample deposited onto a SiO$_2$ coverslip. Maps of the neutral exciton ($\rm X^0$) emission FWHM ($\Gamma_{\rm X^0}$) (b) and integrated intensity $I_{\rm X^0}$ (c). (d) Map of the trion ($\rm X^{\star}$) integrated PL intensity ($I_{\rm X^0}$). No $\rm X^{\star}$ emission is observed over the whole MoSe$_2$/graphene area. Differential reflectance spectra $\Delta R/R$ and PL spectra of (e) MoSe$_2$ and (f) MoSe$_2$/graphene, taken at the spots indicated in white and red on (b-d), respectively. The inset in (f) shows the PL spectrum of MoSe$_2$/graphene on a semi-logarithmic scale. The data are recorded at a temperature of 4~K, in the linear regime using  continuous wave laser excitation at 2.33~eV.}
\label{Fig2}
\end{center}
\end{figure*}

~

\textbf{Bright, single-line emission in TMD/graphene heterostructures} 

Figure~\ref{Fig1} shows the PL spectra of van der Waals heterostructures made from monolayers of  MoS$_2$, MoSe$_2$, WS$_2$ and WSe$_2$ stacked onto graphene monolayers and encapsulated in BN (lower panel). These spectra are compared to those of neighbouring BN-capped TMD regions (upper panel). All TMD/graphene spectra display one single and narrow Lorentzian emission line, with a full width at half maximum (FWHM) of typically 5 meV, suggesting minimal dephasing and disorder (see supplementary Section~1 for the fitting parameters and Section 2 for MoSe$_2$/graphene samples with PL linewidths approaching the lieftime limit). The TMD references also display narrow emission features assigned to $\rm X^0$ (see asterisks in Fig.~\ref{Fig1}, upper panel), but the latter are accompanied with the lower energy emission lines introduced above~\cite{Cadiz2017,Wang2018,Vaclavkova2018,Liu2019} (see supplementary Section~3 for their assignments). The sharp PL lines in TMD/graphene are slightly redshifted (by $\sim 10~\rm{meV}$) with respect to $\rm X^0$ in the TMD references. By measuring the temperature dependent PL spectra and the differential reflectance (DR) spectra of TMD and TMD/graphene, we can unambiguously assign these single lines to $\rm X^0$ (see Fig.~\ref{Fig2} and supplementary Section~4). Therefore, we conclude (i) that no $\rm X^{\star}$ emission is measured in TMD/graphene and (ii) that the $\rm X^0$ redshift in TMD/graphene arises from dielectric screening~\cite{Raja2017,Froehlicher2018}. Crucially, we note that PL quenching of the $\rm X^0$ line is moderate, systematically of less than one order of magnitude in all samples under study (see Fig.~\ref{Fig1} and supplementary Section~1). The difference between PL from TMD and TMD/graphene heterostructures is particularly striking in the case of tungsten-based TMDs. In these dark materials, hot luminescence from $\rm X^0$ is quite inefficient and lower lying emission lines dominate the PL spectra, especially in WS$_2$, (see Fig.~\ref{Fig1} and supplementary Section~3). As discussed below, all these features are much longer lived than $\rm X^0$ in TMD~\cite{Lagarde2014,Robert2016,Robert2017,Nagler2018} and are thus literally washed out in the emission spectra of TMD/graphene heterostructures due to fast non-radiative transfer to graphene.

~

\textbf{Complete TMD neutralization}

We now focus on the  case of MoSe$_2$/graphene heterostructures, with PL mapping, as well as typical DR and PL spectra shown in Fig.~\ref{Fig2} for a BN-capped sample deposited onto a glass coverslip. The PL spectrum of BN-capped MoSe$_2$ is composed of two lines  with similar intensities,  shifted by~28~meV. The high and low-energy line are assigned to $\rm X^0$ and $\rm X^{\star}$ PL~\cite{Ross2013}, respectively. An $\rm X^{\star}$ absorption feature emerges on the DR spectrum, with an amplitude considerably smaller than that of the $\rm X^0$ DR feature (Fig.~\ref{Fig2}e). In contrast, the BN-capped MoSe$_2$/graphene region of the sample displays only $\rm X^0$ absorption and emission features. As shown in Fig.~\ref{Fig2}b-d, these observations can be consistently made over the whole ($>40~\mu\rm m^2$) area of a coupled MoSe$_2$/graphene region. This is an important point, since TMD-based van der Waals heterostructures are known to be spatially inhomogeneous and hence, the observation of ``trion-free'' spectra might be accidental. In return, the absence of $\rm X^{\star}$ emission can be exploited as a reliable probe of the coupling between TMD and graphene. 

We assign the absence of $\rm X^{\star}$ absorption and emission features in TMD/graphene to the transfer of all the native dopants in the TMD (either electrons or holes, with a typical density on the order of $10^{11}-10^{12}~\rm{cm^{-2}}$) to graphene. Such static charge transfer leads to a slight increase of the Fermi level of graphene (typically by less than 100~meV) and to the observation of intrinsic absorption and emission. Alternate scenarii, involving residual doping in the TMD but massive quenching of  $\rm X^{\star}$ formation and/or radiative recombination can safely be ruled out (see supplementary Section~5.1). TMD neutralization is corroborated by room temperature Raman scattering~\cite{Froehlicher2018} and PL measurements (see supplementary Section~5.2). 



\begin{figure*}[!tbh]
\begin{center}
\includegraphics[width=0.97\linewidth]{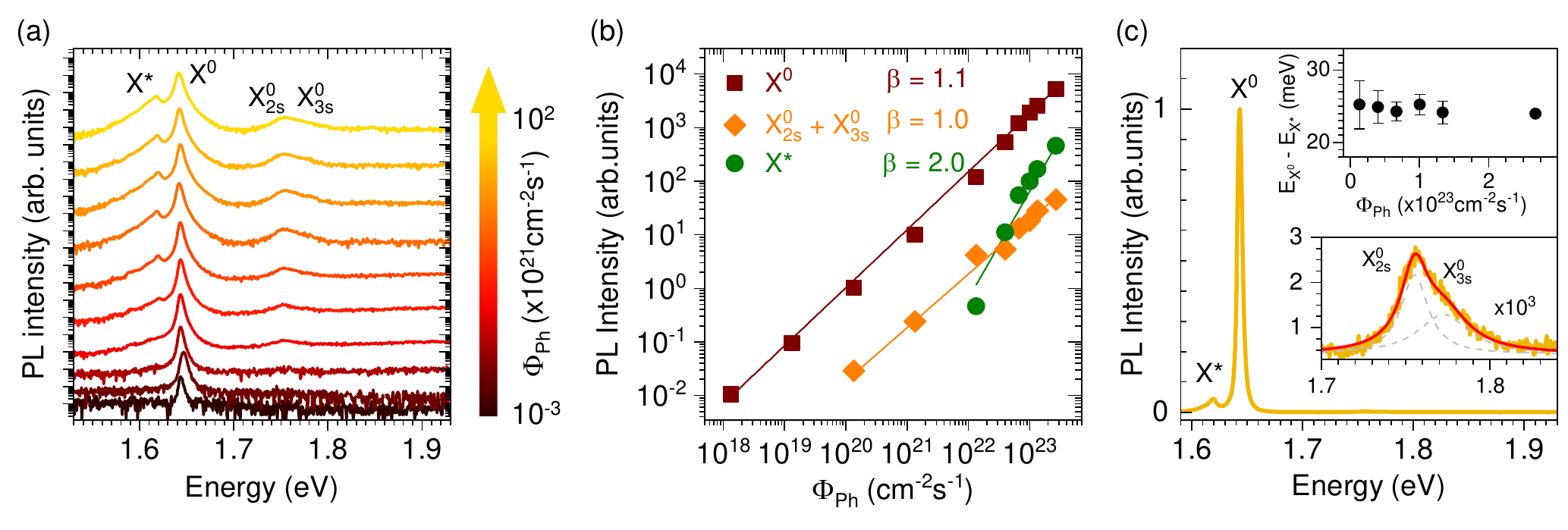}
\caption{\textbf{Laser power dependent photoluminescence in BN-capped TMD/graphene at 4~K.} (a) Semi-logarithmic cascade plot of the PL spectra of the BN-capped MoSe$_2$/graphene sample shown in Fig.~\ref{Fig1},\ref{Fig2}, recorded under cw excitation at 2.33~eV with increasing photon fluxes $\Phi_{\rm{ph}}$. The spectra are normalized by $\Phi_{\rm{ph}}$ and by the integration time and are vertically shifted for clarity. (b) Dependence of the integrated intensity of the neutral 1s exciton ($\rm X^0$ brown), sum of the first and second excited  neutral excitons $\rm X^0_{2\rm s}$ and $\rm X^0_{3\rm s}$ (orange) and photoinduced trion ($\rm X^{\star}$, green).  The solid lines are fits based on a power law and the critical exponents $\beta$ are indicated.  (c) PL spectrum recorded at  $\Phi_{\rm{ph}}=3\times 10^{23}\rm{cm^{-2}\ s^{-1}}$ showing the $\rm X^0$ and $\rm X^{\star}$  features on a linear scale. The $\rm X^0_{2\rm s}$ and $\rm X^0_{3\rm s}$ features are shown in the lower inset, where the red line is a double-Lorentzian fit to the data, with the $\rm X^0_{2\rm s}$ and $\rm X^0_{3\rm s}$ features individually shown with dashed grey lines. The upper inset in (c) shows the energy difference between $\rm X^0$ and $\rm X^{\star}$ as a function of $\Phi_{\rm{ph}}$.}
\label{Fig3}
\end{center}
\end{figure*}

~

\textbf{Photostability and dielectric screening}

\begin{figure*}[!th]
\begin{center}
\includegraphics[width=0.8\linewidth]{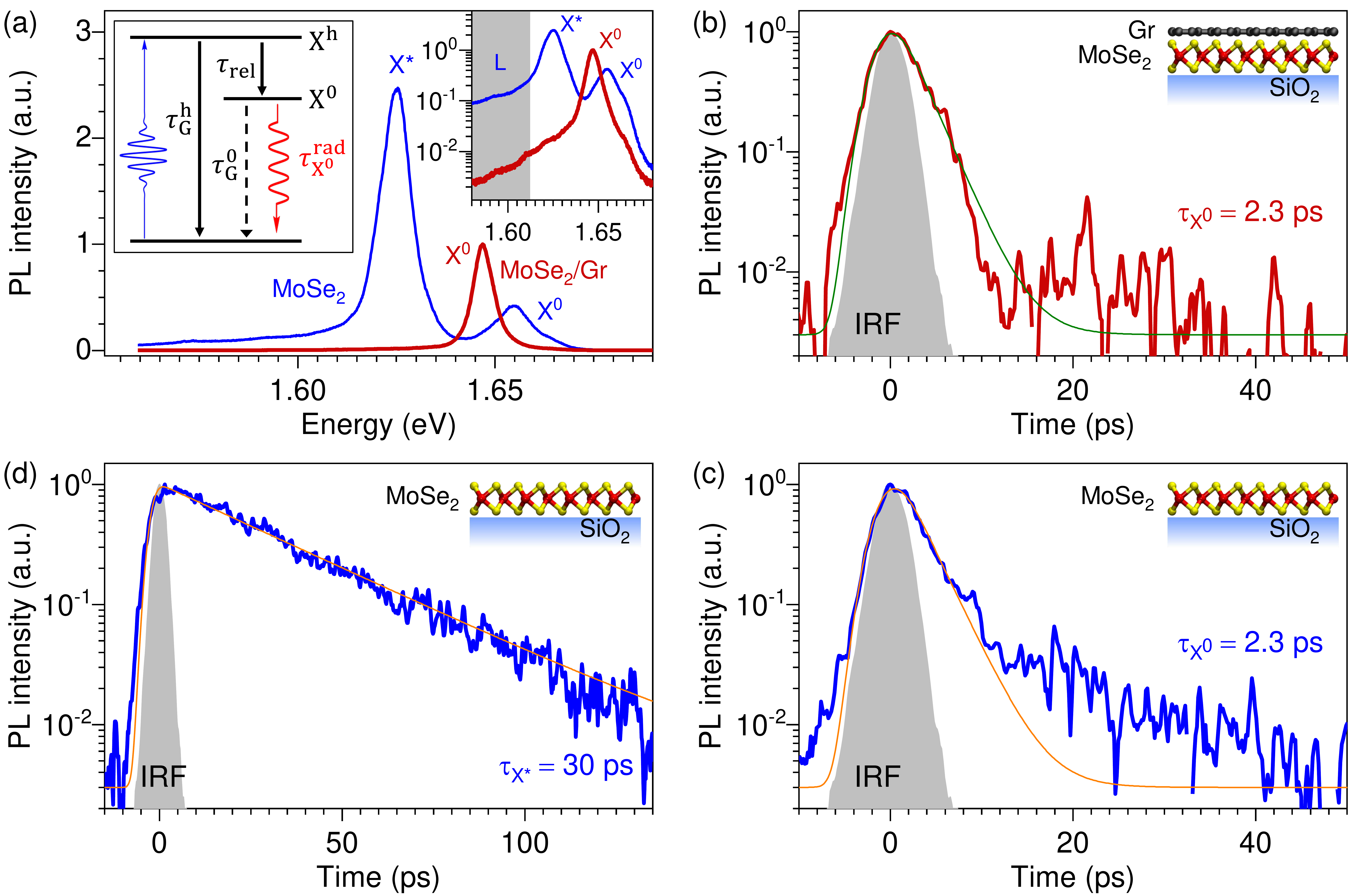}
\caption{\textbf{Low-temperature exciton dynamics.} (a) PL spectra of bare MoSe$_2$ monolayer (blue) and of a MoSe$_2$/graphene heterostructure (red, thicker line), both deposited onto a SiO$_2$ substrate (see sample sketches in (b)-(d). The right inset shows the data on a semi-logarithmic scale. The grey-shaded area highlights a low-energy PL tail from localised (L) states in bare-MoSe$_2$ that is massively quenched in MoSe$_2$/graphene. A three-level system is shown on the left inset. The blue wavy arrow indicates the exciting laser pump pulse that generates a population of hot excitons ($\rm X^{\rm h}$). The latter relax with a lifetime  $\tau_{\rm{rel}}$ to form cold excitons ($\rm X^0$), from which radiative recombination occurs with a lifetime $\tau^{\rm{rad}}_{\rm X^0}$. The presence of graphene induces non-radiative decay of $\rm X^{\rm h}$ and $\rm X^{\rm 0}$ with timescales $\tau^{\rm{h}}_{\rm G}$ and $\tau^{0}_{\rm G}$, respectively. A similar sketch may be drawn for $\rm X^{\star}$ and other excitonic sates.  Time-resolved PL of the neutral exciton feature ($\rm X^0$)  in MoSe$_2$/graphene (b), and in bare MoSe$_2$ (c). The  longer-lived tail in the bare MoSe$_2$ PL decay is attributed to exciton localisation. (d) TRPL of the trion feature ($\rm X^\star$) in MoSe$_2$. All data were recorded at 14~K in the linear regime under excitation with~ps laser pulses at 1.76~eV. The thin solid lines are mono-exponential fits to the data after convolution with the instrument response function (IRF, grey area). The extracted $\rm X^0$ and $\rm X^{\star}$ lifetimes ($\tau_{\rm X^0}$ and $\tau_{\rm X^{\star}}$, respectively)  are indicated.}
\label{FigTRPL}
\end{center}
\end{figure*}

\begin{figure*}[!th]
\begin{center}
\includegraphics[width=0.8\linewidth]{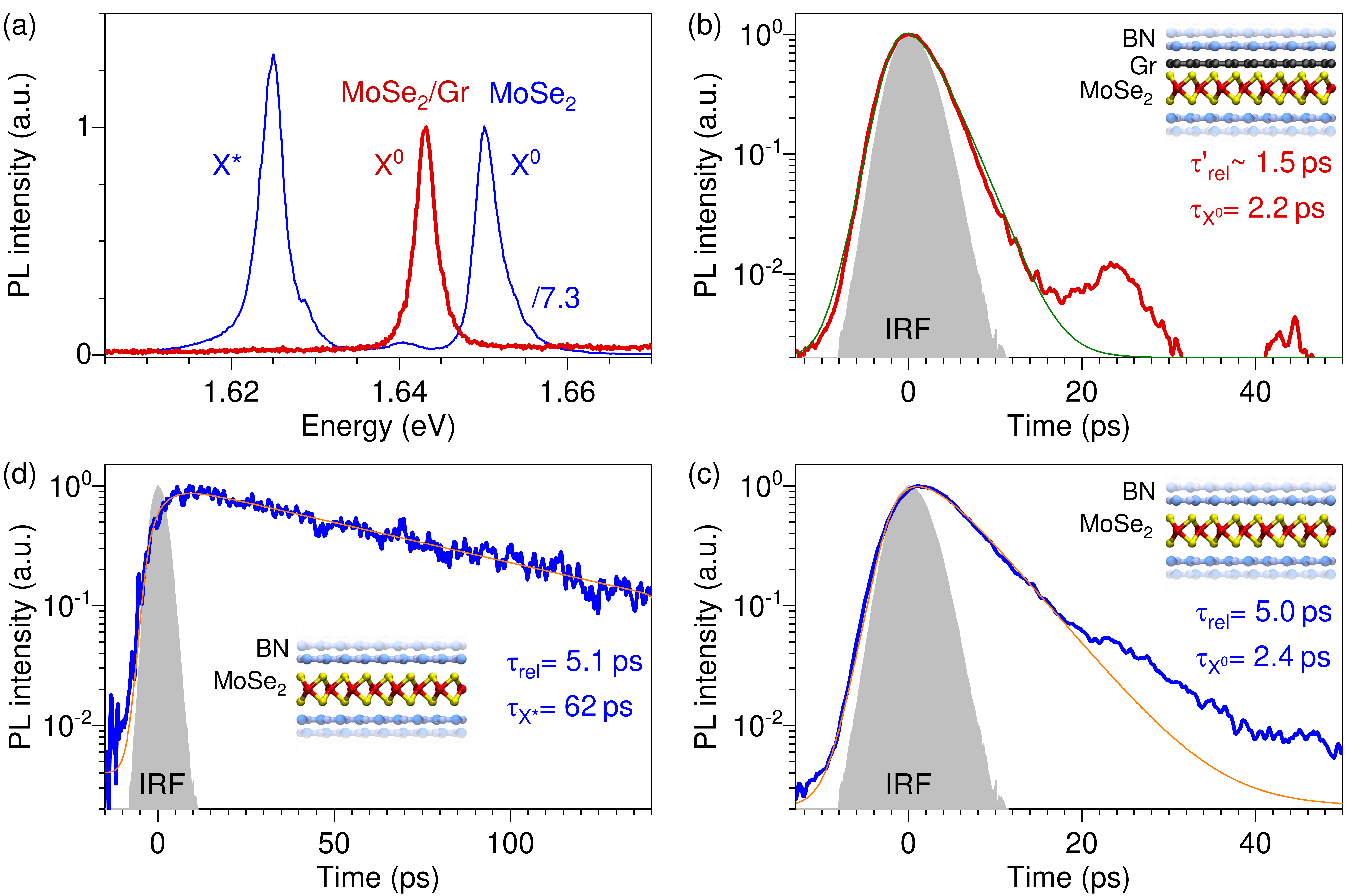}
\caption{\textbf{Evidence for hot exciton transfer to graphene.} (a) PL spectra of BN-capped MoSe$_2$ monolayer (blue) and a BN-capped MoSe$_2$/graphene heterostructure (red), both deposited onto a SiO$_2$ coverslip (see sample sketches in (b)-(d)). The PL spectrum of BN-capped MoSe$_2$ is scaled down by a factor 7.3 for a clearer comparison.  Time-resolved PL of the neutral exciton feature ($\rm X^0$) in BN-capped MoSe$_2$/graphene (b), and in BN-capped MoSe$_2$ (c). The  longer lived tail in the MoSe$_2$ PL decay is attributed to exciton localisation. (d) TRPL of the trion feature ($\rm X^\star$) in MoSe$_2$. All data were recorded at 7~K in the linear regime under excitation with~ps laser pulses at 1.73~eV. The thin solid lines are fits to the data using the three-level model shown in the inset of Fig.~\ref{FigTRPL}a, after convolution with the instrument response function (IRF, grey area). The extracted hot exciton relaxation times ($\tau^{\prime}_{\rm {rel}}$ in (b), $\tau_{\rm {rel}}$ in (c)) as well as the $\rm X^0$ and $\rm X^{\star}$ decay times ($\tau_{\rm X^0}$ and $\tau_{\rm X^{\star}}$, respectively) are indicated.}
\label{FigTRPLE1}
\end{center}
\end{figure*}

To further establish the benefits of coupling TMD to graphene, we show, in Fig.~\ref{Fig3}, the PL spectra of the BN-capped MoSe$_2$/graphene introduced in Fig.~\ref{Fig2} recorded  under continuous wave (cw) photon fluxes (hereafter denoted $\Phi_{\rm{ph}}$) at 2.33~eV, spanning more than five orders of magnitude, from $1\times10^{18}~\rm{cm^{-2}\  \rm s^{-1}}$ to  $3\times10^{23}~\rm{cm^{-2}\ \rm s^{-1}}$, i.e., $\approx 1~\rm {mW}/{\mu \rm m}^2$. Assuming an absorptance of $\sim 10~\%$ at 2.33~eV~\cite{Li2014} and an excitonic lifetime of 2~ps in BN-capped MoSe$_2$/graphene (see Fig.~\ref{FigTRPL},\ref{FigTRPLE1}), we may estimate injected hot exciton densities ranging from $2\times10^{5}~\rm{cm^{-2}}$  up to $6\times10^{10}~\rm{cm^{-2}}$. The $\rm X^{0}$ PL intensity scales quasi-linearly with $\Phi_{\rm{ph}}$. We note, however, the emergence of two faint PL features. The first one is asymmetric, blueshifted by about 120~meV from $\rm X^0$ and its intensity scales linearly with $\Phi_{\rm{ph}}$. We assign this feature to hot luminescence from the neutral excited $\rm X^0_{2\rm s}$ excitons (at $\Delta_{1\rm s-2\rm s}=110~\rm{meV})$ and $\rm X^0_{3\rm s}$ excitons (at $\Delta_{1\rm s-3\rm s}=127~\rm{meV}$). The exciton binding energy is $E_{\rm b}=231~\rm{meV}$ in BN-capped MoSe$_2$, with $\Delta_{1\rm s-2\rm s}\approx168~\rm{meV}$. Assuming for simplicity that  $E_{\rm b}$ scales proportionally to $\Delta_{1\rm s-2\rm s}$, we estimate that the presence of a graphene layer reduces $E_{\rm b}$ down to $\approx~148 ~\rm{meV}$ (see supplementary Section~6). The second feature is redshifted by about $25\pm 2~\rm{meV}$, i.e., a few meV less than the $\rm X^{\star}$ line in MoSe$_2$ (see Fig.~\ref{Fig2}c) and its intensity rises quadratically with $\Phi_{\rm{ph}}$. Therefore, this feature could tentatively be assigned to a biexciton ($\rm{XX}^0$). However, $\rm{XX}^0$ have recently been observed in MoSe$_2$~\cite{Hao2017} and WSe$_2$~\cite{Barbone2018,Ye2018,Paur2019,Li2018} monolayers and display binding energies of $\lesssim 20~\rm{meV}$, significantly lower than the value observed here in a system that undergoes more screening due to the presence of graphene. Thus, the lower energy feature is tentatively assigned to emission from photocreated trions with a slightly reduced binding energy. As previously reported in other low-dimensional materials (e.g.; carbon nanotubes~\cite{Yuma2013}), at sufficiently large exciton densities, bimolecular exciton-exciton annihilation (EEA~\cite{Mouri2014}) can create free carriers. Subsequent photon absorption leads to $\rm X^{\star}$ formation and emission. In this scenario, a quadratic scaling with $\Phi_{\rm {ph}}$ is expected. Noteworthy, $\rm X^0$ emission remains more than one order of magnitude brighter than $\rm X^{\star}$ emission at the highest $\Phi_{\rm{ph}}$ employed here, which justifies why we do not observe a sub-linear rise of the  $\rm X^0$ PL intensity due to EEA. The data in Fig.~\ref{Fig3} also reveals the outstanding photostability of TMD/graphene systems that can sustain photon fluxes above typical PL non-linearity and damage thresholds in bare TMD~\cite{Cadiz2016}, without significant photo-induced heating (see supplementary Sections~5 and~7).

~

\textbf{Exciton dynamics and exciton transfer time}

Let us finally consider exciton dynamics in TMD/graphene. Fig.~\ref{FigTRPL} compares the time-resolved PL (TRPL) of MoSe$_2$ and MoSe$_2$/graphene, recorded at 14~K on another sample deposited on SiO$_2$ and excited below the $\rm X^0_{2\rm s}$ state. As shown in Fig.~\ref{FigTRPL}a, the filtering effect evidenced in BN-capped TMD/graphene (Fig.~\ref{Fig1}) also appears prominently samples supported by a rougher SiO$_2$ substrate (see also supplementary Section~2). A three-level system~\cite{Fang2019} considering hot excitons $\rm X^{\rm h}$, such as finite momentum 1s and possibly 2s excitons, formed shortly after laser excitation~\cite{Steinleitner2017}, $\rm X^0$ (i.e., cold 1s excitons with center of mass momentum within the light cone) and the ground state is shown in the inset of Fig.~\ref{FigTRPL}a. As the PL rise time lies below our time resolution for all measurements on this sample, the TRPL traces in Fig.~\ref{FigTRPL} are simply fit by the convolution of the instrument response function (IRF) with an exponential decay. In keeping with previous reports, the $\rm X^0$ exciton lifetime ($\tau_{\rm X^0}$) is only $\approx 2.3~\rm {ps}$ in bare MoSe$_2$, and can be assigned to the radiative lifetime $\tau^{\rm{rad}}_{\rm X^0}$~\cite{Robert2016}, whereas $\rm X^{\star}$ display a much longer lifetime $\tau_{\rm X^{\star}}=30~\rm{ps}$. Remarkably, within experimental accuracy, $\tau_{\rm X^0}$ has identical values in MoSe$_2$/graphene and in the neighbouring MoSe$_2$ region. This striking result suggests that $\tau_{\rm G}^0$, the non-radiative transfer time of cold excitons ($\rm X^0$) to graphene, is longer than 2~ps. Let us note, that $\tau_{\rm X^0}^{\rm {rad}}$ scales as $E_{\rm b}^{-2}$ and is thus expected to be longer in MoSe$_2$/graphene than in the nearby MoSe$_2$ reference. Hence, observing similar $\tau_{\rm X^0}$ in Fig.~\ref{FigTRPL}b,c may be coincidental and result from the compensation between the increase of $\tau_{\rm X^0}^{\rm {rad}}$ in MoSe$_2$/graphene and non-radiative transfer of $\rm X^0$ to graphene  with an estimated timescale $\tau_{\rm{G}}^0\sim 5 ~\rm{ps}$ (see supplementary Section~8.1). As a result, the $\rm X^0$ emission yield (defined here as the number of emitted photons divided by the number of cold $\rm X^0$) that is near unity in the bare MoSe$_2$ monolayer~\cite{Fang2019} remains high, near $50~\%$ in MoSe$_2$/graphene (see supplementary Section~8.2).  

Still, while the integrated PL intensity from  $\rm X^0$ in MoSe$_2$/graphene on SiO$_2$ is nearly twice that of a close-lying MoSe$_2$ reference (see Fig.~\ref{FigTRPL}a), the \textit{total} PL intensity from MoSe$_2$ remains $\approx 4.5$ times larger than that of MoSe$_2$/graphene. Following generation of $\rm X^{\rm h}$, one may form $\rm X^0$, $\rm X^{\star}$ and localised excitons (e.g., near defects) in bare MoSe$_2$. Localised excitons contribute a low-energy tail to the PL spectra, as shown in Fig.~\ref{FigTRPL}a (shaded area under the blue trace). Conversely, only $\rm X^0$ and localised excitons can be formed in neutral graphene/MoSe$_2$/SiO$_2$, which is consitent with the observed enhancement of $\rm X^0$ emission. Since localised excitons have long lifetimes~\cite{Lagarde2014}, their emission is quenched by graphene, as evidenced in Fig.~\ref{FigTRPL}a (red trace). The sizeable quenching of the \textit{total} PL intensity can tentatively be assigned to fast transfer of $\rm X^{\rm h}$ to graphene on a time scale $\tau_{\rm G}^{\rm h}$ (see inset in Fig.~\ref{FigTRPL}a) that is shorter than $\tau_{\rm{rel}}$, the relaxation time of $\rm X^{\rm h}$ down to the light cone. This scenario is consistent with the fact that the F\"orster-type energy transfer time of TMD excitons to graphene is expected to be maximal for zero momentum excitons (i.e., here $\rm X^0$) and to decrease with exciton momentum~\cite{Basko1999,Selig2019}.
All in all, our results strongly suggest that graphene open up
efficient non-radiative decay pathways that moderately quench $\rm X^{0}$ radiative recombination but instead significantly inhibit $\rm X^0$ formation. Along these lines, we expect that $\rm X^0$ formation should be more efficiently inhibited in BN-capped samples, where $\tau_{\rm {rel}}$ is much longer than in SiO$_2$ supported samples and can be experimentally resolved~\cite{Fang2019}.

~

\textbf{Efficient hot exciton transfer to graphene}

In Fig.~\ref{FigTRPLE1}, we investigate exciton dynamics the BN-capped sample discussed in Fig.~\ref{Fig1}-\ref{Fig3}, following pulsed optical excitation slightly below the $\rm X^0_{2\rm s}$ exciton energy. Importantly, this sample is deposited on a glass coverslip and therefore immune from the cavity effects discussed in ref.~\cite{Fang2019}. As introduced in Fig.~1-3, $\rm X^0$ PL is quenched in MoSe$_2$/graphene, in contrast with the case of the SiO$_2$ supported sample, where $\rm X^0$ PL enhancement is observed (Fig.~\ref{FigTRPL}a). This difference readily suggests that $\rm X^0$ formation is more efficiently quenched in BN-capped TMD/graphene. The TRPL traces are fit by solving the rate equations associated with the three-level system introduced in Fig.~\ref{FigTRPL}a, with two characteristic times for $\rm X^{\rm h}$ relaxation towards $\rm X^0$ (or $\rm X^{\star}$) formation  and  $\rm X^0$ (or $\rm X^{\star}$) decay, respectively. The short and long times correspond to the rise and decay times in the TRPL measurement, respectively~\cite{Fang2019}. Fig.~\ref{FigTRPLE1}b,c reveals that the PL rise time ($\sim 1.5 ~\rm{ps}$) is too close to our resolution limit to be accurately resolved in BN-capped MoSe$_2$/graphene and is $2.4~\rm{ps}$ in the BN-capped MoSe$_2$ reference. The PL decay times are $\approx 2.2~\rm{ps}$ in MoSe$_2$/graphene and $\approx 5~\rm{ps}$ in the MoSe$_2$ reference region, respectively. Interestingly, the $\approx 5.1~\rm{ps}$ PL rise time of the long-lived $\rm X^{\star}$ feature $(\tau_{\rm X^{\star}}=62~\rm{ps})$  in BN-capped MoSe$_2$  is closely matching the $\rm X^0$ decay time. We therefore assign this $\approx 5~\rm{ps}$ time to $\tau_{\rm{rel}}$, the relaxation time of $\rm X^{\rm h}$ down to the $\rm X^{\star}$ and $\rm X^0$ states in BN-capped MoSe$_2$. Hot exciton transfer results in a shortened $\rm X^{\rm h}$ relaxation time $\tau^{\prime}_{\rm{rel}}\lesssim 1.5~\rm{ps}$ in BN-capped MoSe$_2$/graphene.
Finally, the $\rm X^0$ lifetimes $\tau_{\rm X^0}$ in BN-capped MoSe$_2$ and MoSe$_2$/graphene are assigned to the rise and decay time of their respective TRPL traces and are nearly identical ($\approx 2.4~\rm{ps}$ and $\approx 2.2~\rm{ps}$, respectively). With these values, following the same reasoning as for the SiO$_2$-supported MoSe$_2$/graphene sample discussed in Fig.~\ref{FigTRPL}, we can estimate that the $\rm X^0$ emission yield is also near $50~\%$ (see supplementary Section~8.2), demonstrating that non-radiative $\rm X^0$ transfer has similar efficiencies in BN-capped and SiO$_2$-supported MoSe$_2$/graphene heterostructures.
We conclude that a significant part of $\rm X^0$ PL quenching is due to efficient non-radiative transfer of $\rm X^{\rm h}$ enabled by  graphene and that $\rm X^{0}$ PL quenching is more efficient in BN-capped Mose$_2$/graphene than SiO$_2$/MoSe$_2$/graphene due to slower $\rm X^{\rm h}$ relaxation (or equivalently $\rm X^0$ formation) in BN-capped MoSe$_2$.

~

\textbf{Conclusions and outlook}

In closing, we have shown that graphene neutralizes atomically-thin TMDs, leading to the absence of light emission from charged excitonic species. Graphene also enables picosecond non-radiative transfer of TMD excitonic species. Starting from a given initial $\rm X^{\rm h}$ density, transfer of $\rm X^{\rm h}$ to graphene will reduce the maximum achievable density of $\rm X^0$ and of other longer lived neutral excitonic compounds. The latter will be strongly quenched by graphene, whereas the former will be minimally affected owing to their picosecond radiative lifetime. As a result, $\rm X^{0}$ exclusively contribute to the PL spectra of TMD/graphene heterostructures. The measured $\rm X^{0}$ PL intensity is largely determined by the competition between $\rm X^{0}$  formation and non-radiative $\rm X^{\rm h}$ transfer to graphene rather than by the trade-off between radiative $\rm X^{0}$ recombination and non-radiative $\rm X^{0}$ transfer to graphene. 

Graphene is here introduced as narrow-line filter that is naturally tuned to a broad range of emitted photon energies spanning the mid-infrared to the ultraviolet regions. Our two-dimensional design naturally outperforms alternate solutions based on spectrally narrow interference filter that are considerably bulkier and lack tunability (see supplementary Section~9). Going further, high-speed (up to $\sim 1\rm{THz}$), photonic and opto-electronic devices with one single bright and nearly lifetime-limited emission line (see supplementary Section~2) can be envisioned  using TMD/graphene heterostructures. Such devices may also benefit from the excellent electrical contact, photodetection~\cite{Massicotte2016,Arp2019}, electron and spin transport~\cite{Luo2017,Avsar2017} capabilities offered by TMD/graphene heterostructures. One may also foresee progress in cavity quantum electrodynamics~\cite{Schneider2018}, chiral optics~\cite{Chervy2018} and opto-valleytronics~\cite{Mak2018} by jointly exploiting the simple emission spectra of TMD/graphene heterostructures and their record-high degrees of valley coherence and valley polarization of up to 60~$\%$ and 50~$\%$, respectively~\cite{Lorchat2018}.

~

\textbf{{Acknowledgements}} 

The authors thank D. Basko, T. Galvani, L. Wirtz, G. Schull, S. Azzini, T. Chervy, C. Genet, M.A. Semina and M.M. Glazov for fruitful discussions. We are grateful to H. Majjad and M. Rastei for help with AFM measurements, to M. Romeo, F. Chevrier, M. Acosta, A. Boulard and the StNano clean room staff for technical support. We acknowledge financial support from the Agence Nationale de la Recherche (under grants H2DH ANR-15-CE24-0016, 2D-POEM ANR-18-ERC1-0009, D-vdW-Spin,  VallEx, and  MagicValley), from the LabEx NIE (Under Grant ANR-11-LABX-0058-NIE) and from the EUR NanoX (under grant VWspin and MILO). 

~

\textbf{{Competing  interests}}

The authors declare no competing interests.

~

\textbf{{Author contributions}}

S.B. conceived and lead the project, with C.R., D.L. and X.M. supervising the time-resolved PL measurements. E.L. and L.E.P.L. fabricated the samples. E.L., L.E.P.L., C.R., D.L. and S.B. carried out the measurements. E.L., L.E.P.L. and S.B. analysed the data with input from G.F., C.R., D.L. and X.M. T.T. and K.W. provided high-quality hexagonal BN crystals. S.B. wrote the manuscript with input from X.M., C.R., E.L. and L.E.P.L.

~



~

\textbf{{Methods}} 

Our model system is a van der Waals heterostructure formed by stacking a monolayer of graphene onto a TMD monolayer using standard methods as in ref.~\cite{Castellanos2014,Zomer2014}. In this work, we have investigated MoSe$_2$, MoS$_2$, WSe$_2$ and WS$_2$-based heterostructures encapsulated in hexagonal boron nitride (BN, see Fig.~\ref{Fig1}-~\ref{Fig3} and Fig.~\ref{FigTRPLE1}) or  directly deposited on SiO$_2$ substrates (Fig.~\ref{FigTRPL}). All materials were mechanically exfoliated from bulk crystals. Graphene and TMD monolayers were unambiguously identified using room-temperature Raman and PL spectroscopies, respectively. Our samples were investigated at variable temperature (4~K -- 300~K) by means of micro-PL and differential reflectance (DR) spectroscopy using home-built setups. Time-resolved PL measurements were performed on MoSe$_2$-based samples, using a Ti:Sa oscillator delivering $\approx 2~\rm{ps}$ pulses with a repetition rate of 80~MHz and a synchro-scan streak camera with a temporal resolution of $\approx 1.5~\rm {ps}$. All comparisons between results obtained on TMD and on TMD/graphene are based on measurements performed in the same experimental conditions on a same TMD flake partially covered with graphene (see for example Fig.~\ref{Fig2}).

~



~


%
%
%


\onecolumngrid
\newpage
\begin{center}
{\Large\textbf{Supplementary Information for: \\ Dynamically-enhanced strain in atomically-thin resonators}}
\end{center}

\setcounter{equation}{0}
\setcounter{figure}{0}
\setcounter{section}{0}
\renewcommand{\thetable}{S\arabic{table}}
\renewcommand{\theequation}{S\arabic{equation}}
\renewcommand{\thefigure}{S\arabic{figure}}
\renewcommand{\thesection}{S\arabic{section}}
\renewcommand{\thesubsection}{S\arabic{section}\alph{subsection}}
\renewcommand{\thesubsubsection}{S\arabic{section}\alph{subsection}\arabic{subsubsection}}

\linespread{1.4}\selectfont

\bigskip

\tableofcontents

\clearpage

\textbf{\textsc{Note on the labelling of excitonic states\\}}
\noindent In the main text, $\rm X^0$ refers to bright, neutral 1s A exciton with center of mass momentum within the light cone. In the discussion associated with Fig.~4 and Fig.~5, these excitons are referred to as \textit{cold} excitons, as opposed to \textit{hot} excitons ($\rm X^{\rm h}$), i.e., 1s excitons with center of mass momentum outside the light cone. In the following sections, we shall consider \textit{excited} states from the A excitonic manifold, labelled $\rm X^0_{2\rm s}$, $\rm X^0_{3\rm s}$, as well as the 1s B exciton $\rm B_{1\rm s}$. When necessary, we will use the full $\rm X^0_{1\rm s}$ notation to refer to cold excitons.

\section{PL peak energies, linewidths and quenching factors }

\setlength{\tabcolsep}{0.5cm} 

\begin{table}[h!] 
\begin{center} 
\begin{tabular}{c c c c c} 
\hline \hline\\[-3.5ex] 
& $E_{\rm X^0}~\rm{(eV)}$ & $\Gamma_{\rm X^0}~\rm{(meV)}$ & $Q_{\rm X^0}=\frac{I_{\rm X^0}}{I^{\prime}_{\rm X^{0}}}$ & $Q_{\rm {tot}}=\frac{I_{\rm {tot}}}{I^{\prime}_{\rm X^{0}}}$ \\[1.5ex] 
\hline 
MoS$_2$&1.933&4.0 & \multirow{2}{*}{3.9} & \multirow{2}{*}{4.3}\\[0.5ex]
MoS$_2$/Gr&1.918&3.5  \\[0.5ex]
\hline
MoSe$_2$&1.649&5.6 &  \multirow{2}{*}{7.1} & \multirow{2}{*}{13.6} \\[0.5ex]
MoSe$_2$/Gr&1.635&4.4  \\[0.5ex]
\hline
MoSe$_2$&$1.655^{\star}$&$11.5^{\star}$ & \multirow{2}{*}{$0.5^{\star}$} & \multirow{2}{*}{$4.5^{\star}$} \\[0.5ex]
MoSe$_2$/Gr&$1.647^{\star}$&$5.7^{\star}$  \\[0.5ex]
\hline
WS$_2$&2.084&9.7 &  \multirow{2}{*}{3.0} & \multirow{2}{*}{227}   \\[0.5ex]
WS$_2$/Gr&2.067&4.3  \\[0.5ex]
\hline
WSe$_2$&1.728&8.7 & \multirow{2}{*}{3.7} & \multirow{2}{*}{21}  \\[0.5ex]
WSe$_2$/Gr&1.721&7.2 \\[0.5ex]
\hline \hline
\end{tabular}
\caption{\textbf{Emission characteristics for the samples shown in Fig.~1 of the main text.} Peak emission energy $(E_{\rm X^0})$, full width at half maximum of the $\rm X^0$ emission line (FWHM, denoted $\Gamma_{\rm X^0}$) for the PL spectra of the four BN-capped samples discussed in Fig.~1 in the main text, as well as for the sample deposited on SiO$_2$ discussed in Fig.~4 of the main text (data highlighted with $\star$ symbols). We also consider two PL quenching factors: first, $Q_{\rm X^0}$ the ratio between the integrated intensities of the $\rm X^0$ line in TMD ($I_{\rm X^0}$) and in TMD/graphene ($I^{\prime}_{\rm X^{0}}$); second, $Q_{\rm {tot}}$ the ratio between the total PL integrated intensity in TMD $(I_{\rm {tot}})$ and $I^{\prime}_{\rm X^{0}}$ (that corresponds to the total PL intensity in TMD/graphene). The data were extracted from Lorentzian fits for BN-capped samples and graphene/MoSe$_2$/SiO$_2$ and from a Voigt fit for vacuum/MoSe$_2$/SiO$_2$. Noteworthy, we observe that $\Gamma_{\rm X^0}$ is generally smaller on TMD/graphene heterostructures than on TMD, suggesting that graphene further reduces disorder induced broadening and pure dephasing (see also Sec.~\ref{SecGamma}). This effect is particularly striking in our SiO$_2$ supported sample, since the bare MoSe$_2$ reference does not benefit from BN encapsulation. $Q_{\rm X^0}$ remains below 10 in all samples under study and is even smaller than 1 in our SiO$_2$ supported sample.}
\label{T1}
\end{center}
\end{table}

\clearpage

\section{Approaching the homogeneous limit}
\label{SecGamma}
As shown in Table~\ref{T1}, most of our BN-capped samples display $\Gamma_{\rm X^0}$ of a few meV and, in any event, significantly below 10~meV (see Table~\ref{T1}). Considering that $\tau_{\rm X^{0}} \approx 2~\rm{ps}$ (see Fig.~4,5 and also Sec.~\ref{SecTRPL}), we would expect a lifetime-limited $\Gamma_{\rm X^{0}}^{\rm{min}}=\hbar/\tau_{\rm X^{0}}\approx 0.3~\rm{meV}$.  

\begin{figure*}[!tbh]
\begin{center}
\includegraphics[width=0.55\linewidth]{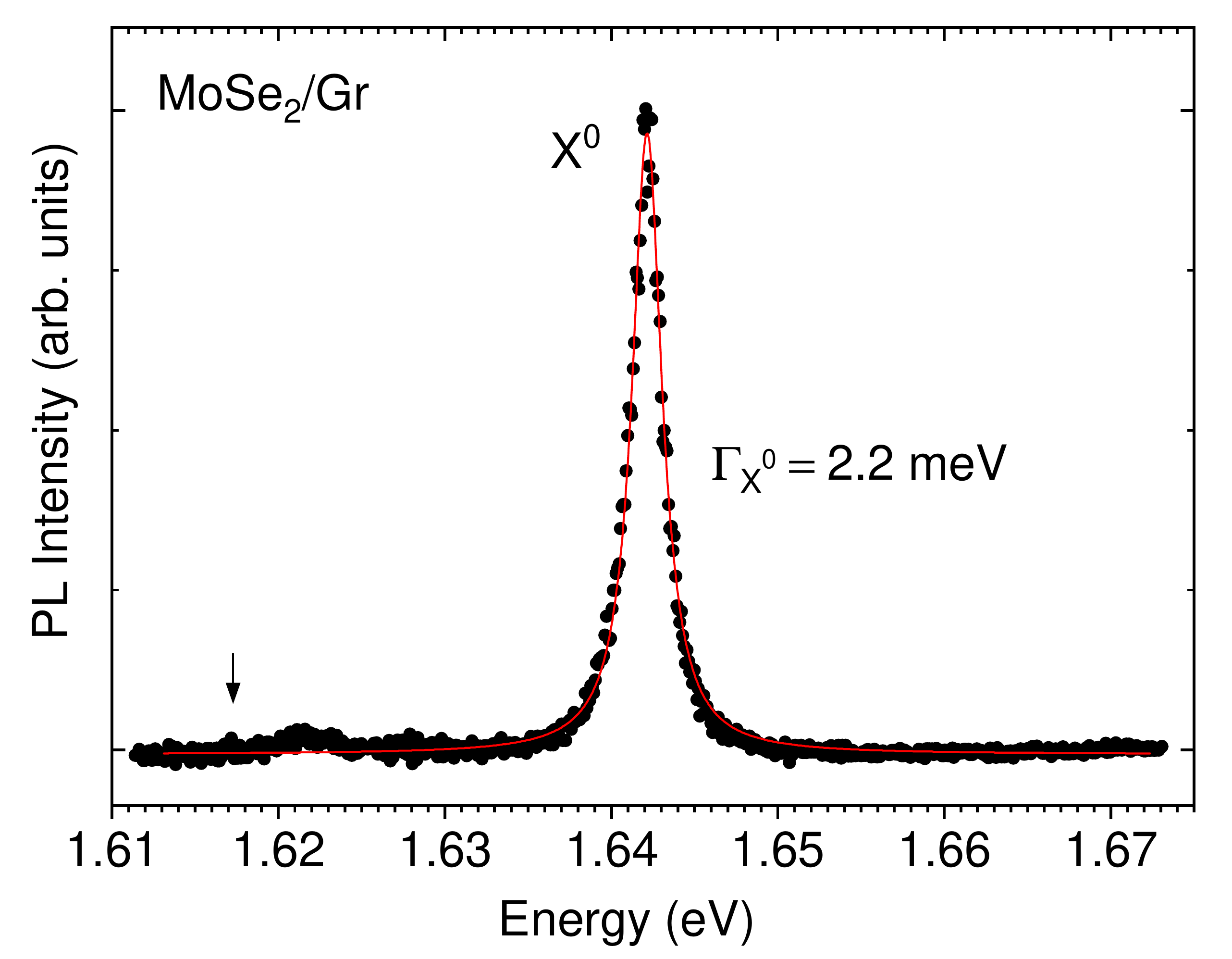}
\caption{\textbf{Narrow PL linewdidth in BN-capped MoSe$_2$/graphene at low temperature.} PL spectrum of another BN-capped MoSe$_2$/graphene sample exhibiting a narrow $\Gamma_{\rm X^0}=2.2~\rm{meV}$. The spectrum (black circles) was recorded at $T=4~\rm K$ with a laser photon energy of 2.33~eV and fit to a Lorentzian function (red line). The black arrow at 1.617~eV indicates the expected location of the $\rm X^{\star}$ line that is not detectable.}
\label{FigS_iBNF2}
\end{center}
\end{figure*}

Figure~\ref{FigS_iBNF2} shows the PL spectra of another BN-capped MoSe$_2$/graphene sample exhibiting a particularly narrow $\Gamma_{\rm X^0}$ down to 2.2~meV. Although  $\Gamma_{\rm X^{0}}$ remains dominated by broadening due to disorder and dephasing, these results show that one can approach the homogeneous limit in TMD/graphene heterostructures. 

\begin{figure*}[!tbh]
\begin{center}
\includegraphics[width=0.65\linewidth]{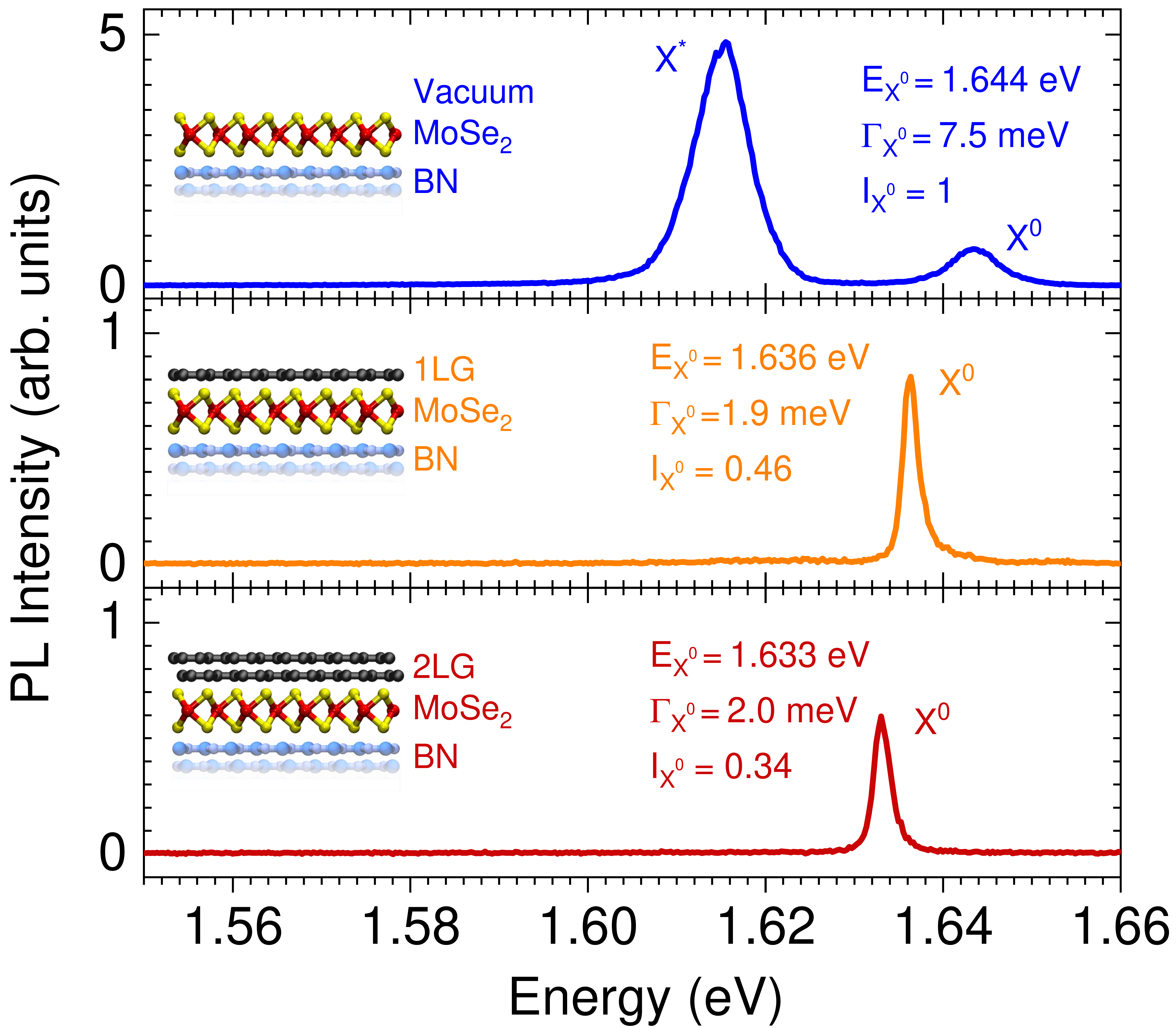}
\caption{\textbf{Bright, single and narrow line emission in BN/MoSe$_2$/graphene.} PL spectra of a MoSe$_2$ monolayer directly deposited on BN (top, blue line), partly covered by a single layer of graphene (middle, orange) and by a bilayer of graphene (bottom, dark red). The $\rm X^0$ exciton energy, linewidth and integrated PL intensity ($I_{\rm X^0}$, in arbitrary units) are indicated. The spectra were recorded at $T=15~\rm K$ with a laser photon energy of 2.33~eV.}
\label{FigS_E2018_comp}
\end{center}
\end{figure*}

Along this line, we observed that narrow PL linewidth can also be achieved in MoSe$_2$/graphene heterostructures without the need for an extra BN top layer. In Fig.~\ref{FigS_E2018_comp}, we report PL spectra of a MoSe$_2$ monolayer deposited onto boron nitride (BN) and covered only by single (1LG) and bilayer (2LG) graphene. We observe nearly identical and narrow $\Gamma_{\rm X^0}$, near 2~meV (FWHM) in the 1LG/MoSe$_2$/BN and 2LG/MoSe$_2$/BN heterostructures. The $\rm X^0$ PL intensity $I_{\rm X^0}$ is only quenched by a factor 2.2 in 1LG/MoSe$_2$/BN as compared to the vacuum/MoSe$_2$/BN reference and, expectedly~\cite{Chen2010} the $\rm X^0$ PL quenching factor $Q_{\rm X^0}$ (see Table~\ref{T1}) only increases up to a factor of 3 in 2LG/MoSe$_2$/BN. As we will further discuss in Sec.~\ref{Metal}, these results demonstrate that using a single layer of graphene is an optimal solution since $\rm X^0$ emission is already very efficiently filtered with minimal PL quenching (see also Table~\ref{T1} and Fig.~4 in the main text for related data on SiO$_2$ supported MoSe$_2$/graphene).

\clearpage

\section{Assigning the low-energy emission lines in WSe$_2$ and W\element{S}$_2$}

\begin{figure*}[!tbh]
\begin{center}
\includegraphics[width=0.8\linewidth]{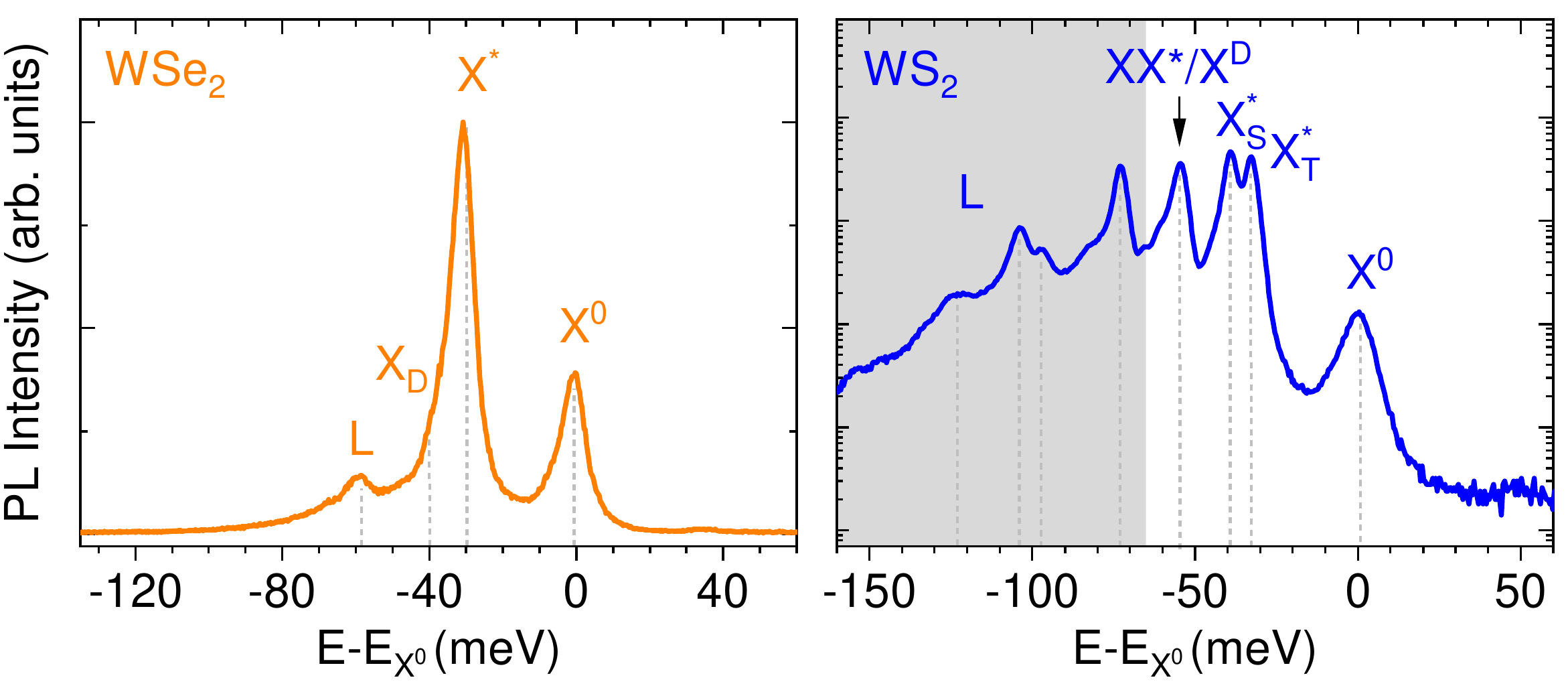}
\caption{\textbf{Photoluminescence lines in WSe$_2$ and WS$_2$.} PL spectra of BN-capped WSe$_2$ (left) and WS$_2$ (right) recorded at 14~K under cw laser excitation at 2.33~eV and 1.96~eV, respectively, as in Fig.~1 of the main text. The origin of the energy axis is taken at the $\rm X^0$ peak energy. The PL spectrum of  WS$_2$ is shown on a semi-logarithmic scale for clarity. The low-energy PL lines below $\rm X^0$ are assigned in keeping with recent reports~\cite{Wang2017,Courtade2017,Vaclavkova2018}. $\rm X^{\rm D}$ denotes a spin-dark exciton, $\rm X^{\star}_{\rm S}$ and $\rm X^{\star}_{\rm T}$ are the singlet and triplet trions, $\rm{XX}^{\star}$ denotes a charged biexciton and $\rm L$ refers to localised states, whose physical origin is still debated. Let us note that exciton-phonon replicas may also contribute to the low energy side of the PL spectra~\cite{Liu2019}. Also, in the case of WS$_2$, $\rm{XX}^{\star}$ and $\rm{X}^{\rm D}$ are expected at very close energies~\cite{Nagler2018,Wang2017}.}
\label{FigS_WS2_WSe2}
\end{center}
\end{figure*}

\clearpage

\section{Identifying the $\rm X^0$ line in TMD/graphene heterostructures}

\begin{figure*}[!tbh]
\begin{center}
\includegraphics[width=0.85\linewidth]{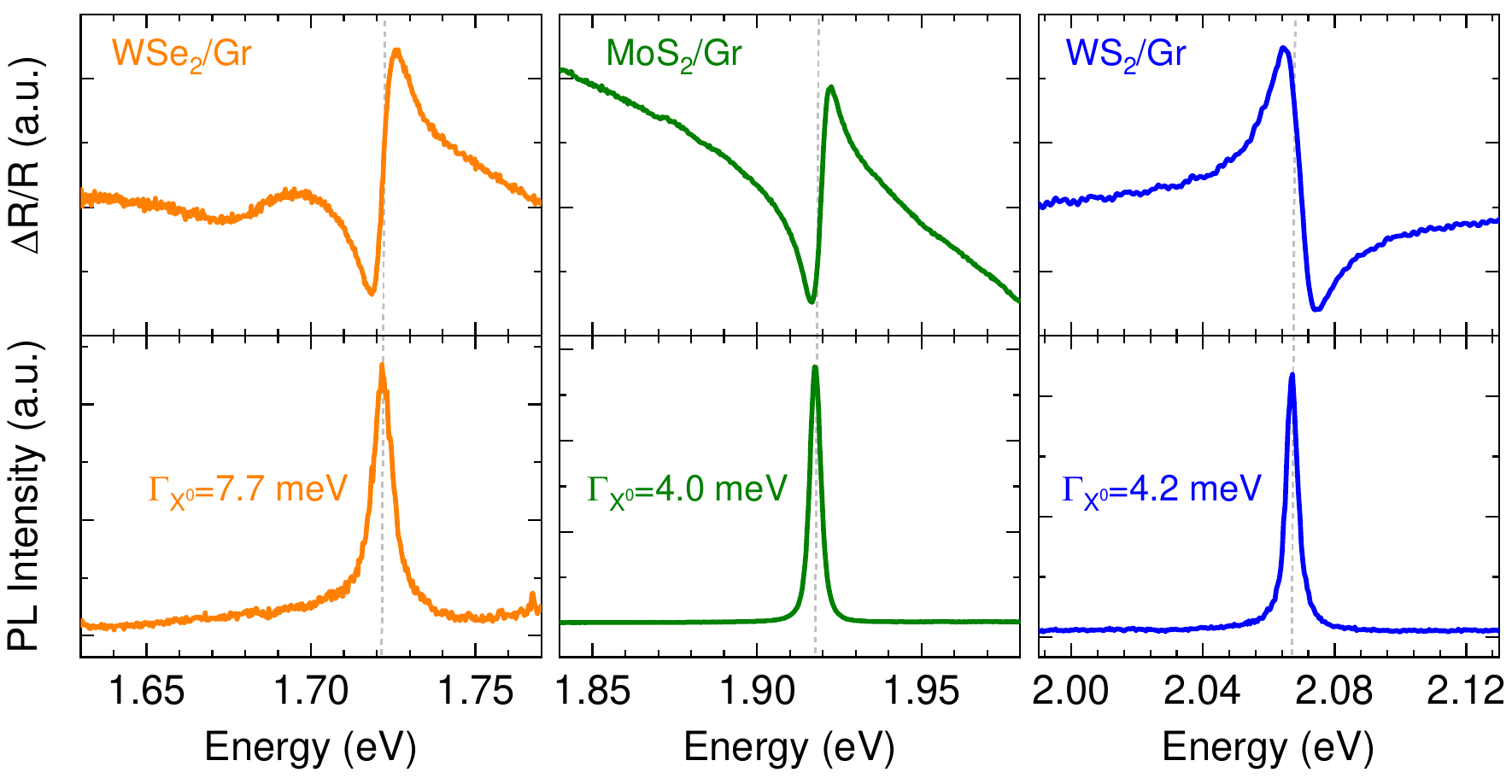}
\caption{\textbf{Identifying the $\rm X^0$ line.} Differential reflectance (DR) and PL spectra of BN-capped WSe$_2$/graphene (left), MoS$_2$/graphene (centre), and WS$_2$/graphene (right) heterostructures, akin to the measurements in Fig.~2f of the main text. The prominent DR and PL features are assigned to $\rm X^0$. All data were recorded below 20~K in the linear regime under continuous wave (cw) laser excitation at 2.33~eV (MoS$_2$/graphene, WS$_2$/graphene) and 1.96~eV (WSe$_2$/graphene). The full width at half maximum of the PL spectra $\Gamma_{\rm X^0}$ are indicated. Note that the data on WSe$_2$ stem another sample than the one shown in Fig.~1 in the main text. The broad DR feature observed near 1.68 eV in this sample is an artifact arising from the underlying substrate.}
\label{FigS_DR}
\end{center}
\end{figure*}

\begin{figure*}[!th]
\begin{center}
\includegraphics[width=1\linewidth]{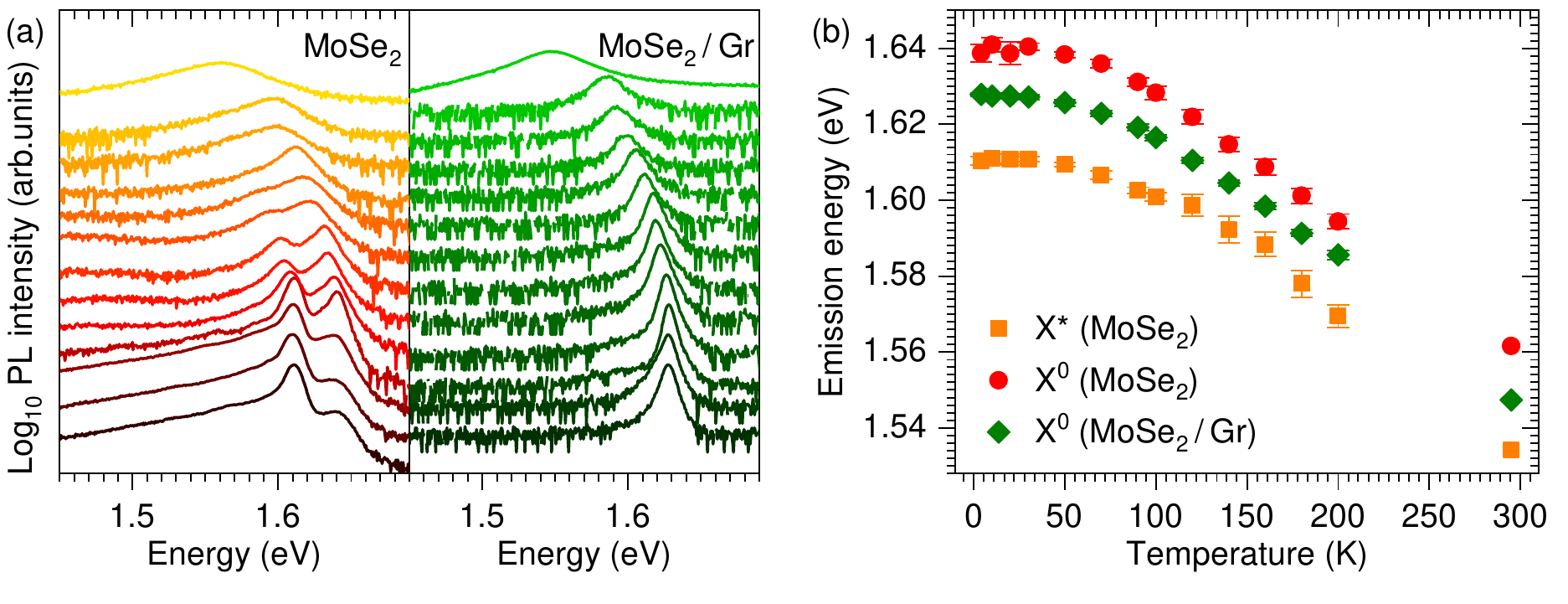}
\caption{\textbf{Temperature dependent PL spectroscopy of the SiO$_2$ supported sample discussed in Fig.~4 of the main text.} (a) Cascade plot of the PL spectra (shown in semi-logarithmic scale) of MoS$2$ and MoSe$_2$/graphene  regions. (b) Temperature dependent peak energy of the MoSe$_2$ exciton ($\rm X^0$, red circles) and trion ($\rm X^{\star}$, orange squares) lines. The prominent emission line in the MoSe$_2$/graphene region (green diamonds) lies between these two lines and is assigned to $\rm X^0$, as also discussed in the main text. The slight ($\sim 10~\rm{meV}$) redshift of the $\rm X^0$ line is due to dielectric screening as discussed in Ref.~\onlinecite{Raja2017,Froehlicher2018}. All data were recorded in the linear regime under continuous wave (cw) laser excitation at 2.33~eV.}
\label{FigS1}
\end{center}
\end{figure*}

\begin{figure*}[!tbh]
\begin{center}
\includegraphics[width=0.5\linewidth]{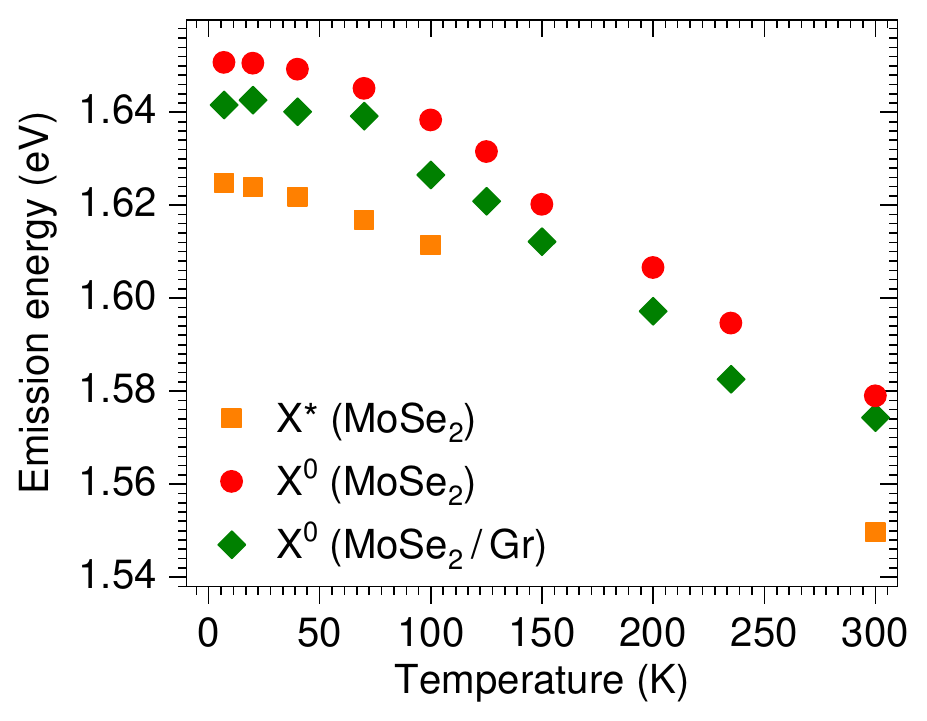}
\caption{\textbf{Temperature dependent PL spectroscopy of the BN-capped MoSe$_2$/graphene sample discussed in Fig.~1-3 and 5 of the main text.} Temperature dependent peak energy of the MoSe$_2$ exciton ($\rm X^0$, red circles) and trion ($\rm X^{\star}$, orange squares) lines. The prominent emission line in the MoSe$_2$/graphene region (green diamonds) lies between these two lines and is assigned to $\rm X^0$, as also discussed in the main text. The  slight ($\sim 10~\rm{meV}$) redshift of the $\rm X^0$ line is due to dielectric screening as discussed in Ref.~\onlinecite{Raja2017,Froehlicher2018}. All data were recorded in the linear regime under continuous wave (cw) laser excitation at 2.33~eV.}
\label{FigS2}
\end{center}
\end{figure*}


\clearpage

\section{Supplementary discussion on the absence of PL from trions}

\subsection{Neutralizing a TMD monolayer with graphene}

\begin{figure*}[!h]
\begin{center}
\includegraphics[width=0.69\linewidth]{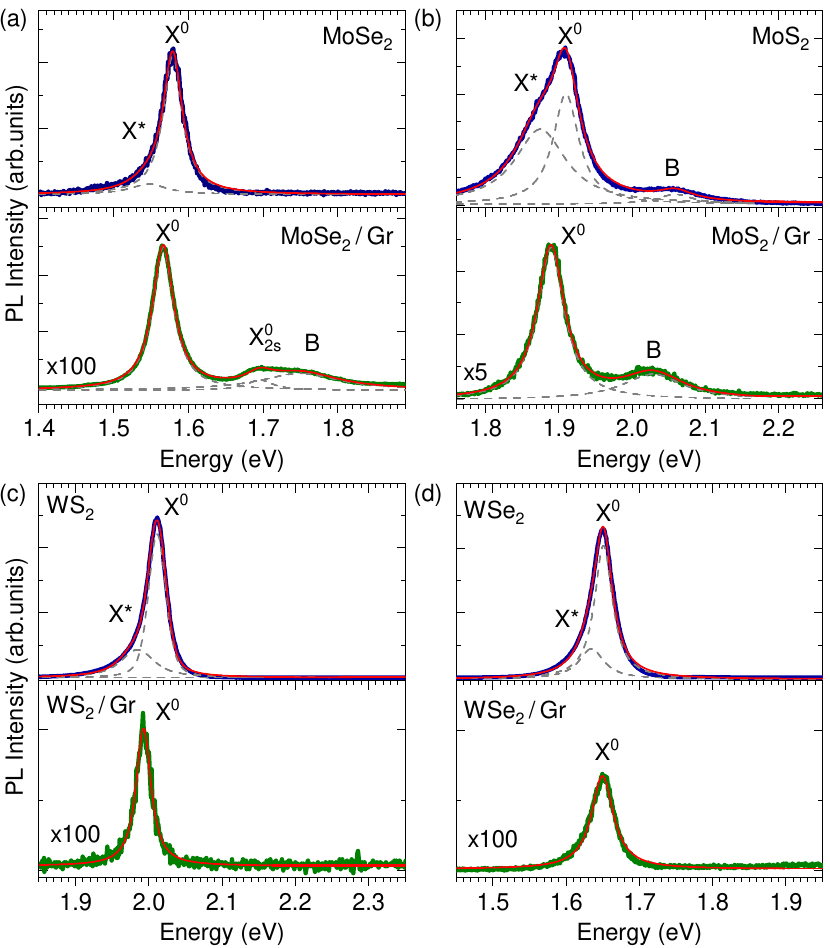}
\caption{\textbf{Trion-free PL spectra at room temperature.} PL spectra of BN-capped TMD/graphene heterostructures compared to those of a nearby BN-capped TMD reference, all recorded in ambient air in the linear regime under cw laser excitation at 2.33~eV. The PL spectra in TMD-graphene are quasi symmetric whereas the PL spectra from TMD exhibit a lower energy shoulder that arises predominantly from trions $(\rm X^{\star})$. The scaling factors allow estimating the large room temperature PL quenching factors~\cite{Froehlicher2018} that strongly contrast with the low $\rm X^0$ PL quenching factors observed at cryogenic temperatures (see main text and Table~\ref{T1}). The $\rm X^0$ lines are slightly redshifted in TMD/graphene, as discussed in the main text, in Fig.~\ref{FigS1},~\ref{FigS2} and in Ref.~\onlinecite{Raja2017,Froehlicher2018}. The red lines are multi-Lorentzian fits to the data, with their different components shown with grey dashed lines. Hot luminescence from excited excitonic states (e.g., $\rm X^0_{2\rm s}$ and B excitons) is clearly visible in MoS$_2$/graphene and MoSe$_2$/graphene.}
\label{FigS3}
\end{center}
\end{figure*}

\begin{figure*}[!tbh]
\begin{center}
\includegraphics[width=0.75\linewidth]{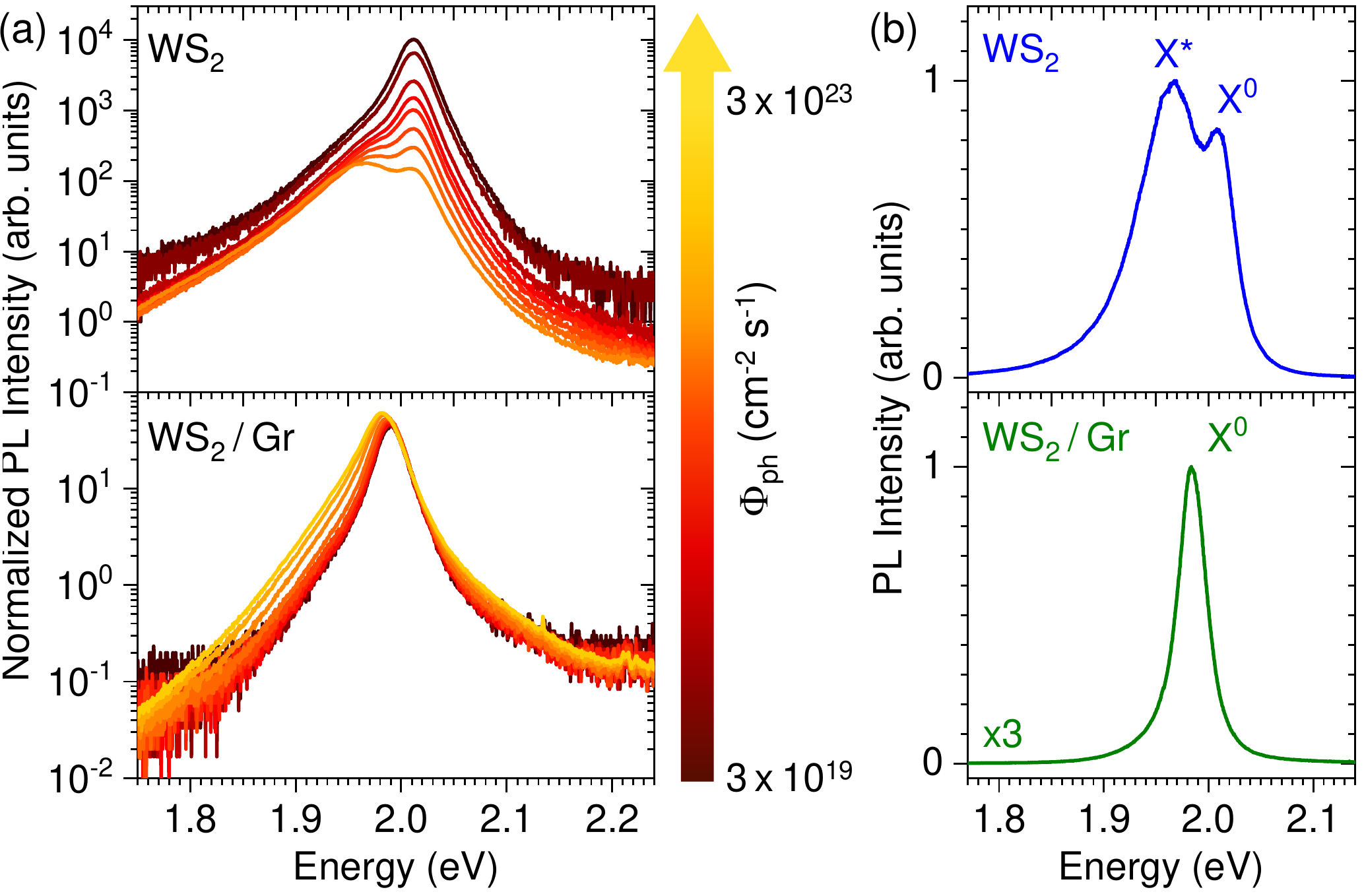}
\caption{\textbf{Neutrality and photostability under high photon flux at room temperature.} (a) Laser power dependent photoluminescence spectra of a BN-capped WS$_2$/graphene heterostructure compared to a nearby BN-capped WS$_2$ reference, recorded in ambient air using cw laser excitation at 2.33~eV. The spectra are shown on a semi-logarithmic scale and are normalised by the incoming photon flux ($\Phi_{\rm ph}$) and the integration time. $\Phi_{\rm ph}$ is color-coded with a gradient ranging from dark red (low $\Phi_{\rm ph}\sim 100~\rm{nW}/\mu\rm m^2$) to yellow (high $\Phi_{\rm ph}\sim 1~\rm{mW}/\mu\rm m^2$). PL saturation due to exciton-exciton annihilation~\cite{Mouri2014} is clearly visible in WS$_2$, whereas a quasi linear scaling is observed in WS$_2$/graphene.  The PL spectra in WS$_2$ remain quasi symmetric even under high $\Phi_{\rm ph}$, whereas the PL spectra from the TMD reference exhibit a lower energy shoulder, assigned to trion ($\rm X^{\star}$) emission. The latter grows significantly as $\Phi_{\rm ph}$ increases and ultimately overcomes the $\rm X^0$ line, as shown in (b) on the selected spectra recorded at $\Phi_{\rm ph}\approx 2\times 10^{23}\rm{cm^{-2}s^{-1}}$ and plotted on a linear scale.}
\label{FigS4}
\end{center}
\end{figure*}

Different scenarii may explain the absence of $\rm X^{\star}$ absorption and emission features in TMD/graphene. Let us first assume that the TMD coupled to graphene remains doped. Trions may then form rapidly but their slow radiative recombination (in the 20-200~ps range~\cite{Robert2016,Fang2019}) is quenched by graphene. We may thus still be able to observe an $\rm X^{\star}$ absorption feature in the DR spectra, in contrast with the measurements in Fig.~2e. Furthermore, since the $\rm X^{\star}$ PL feature is more than three orders of magnitude weaker than the  $\rm X^{0}$ PL feature (see inset in Fig.~2f), the $\rm X^{\star}$ lifetime would be reduced to less than $\sim 100~\rm{fs}$, an unrealistically low value.  Alternatively, graphene may fully quench hot excitons before they can relax and form $\rm X^0$ and $\rm X^{\star}$ near resident carriers. However, $\rm X^{\star}$ and  $\rm X^0$ formation are equally fast in MoSe$_2$ (see Ref.~\cite{Robert2016,Fang2019} and Fig.~4,5). As a result, this scenario would imply massive $\rm X^{0}$ PL quenching, in obvious contradiction with our observations. The third and most plausible course of events is that all the native dopants in the TMD (either electrons or holes, with a typical density on the order of $10^{11}-10^{12}~\rm{cm^{-2}}$) transfer to graphene, leading to a slight increase of the Fermi level of graphene (typically by less than 100~meV) and to the observation of intrinsic absorption and emission. This scenario is corroborated by room temperature Raman scattering~\cite{Froehlicher2018} and PL measurements (see Fig.~\ref{FigS3},~\ref{FigS4}). 

\clearpage

\subsection{Evidence for TMD neutralization at room temperature.}

\clearpage

\section{Exciton binding energy}
\label{Eb}
Encapsulation in BN and more generally screening by a surrounding medium significantly reduce the exciton binding energy  $E_{\rm b}$, as discussed for instance in Ref.~\onlinecite{Raja2017}. In this section, we discuss PL and differential reflectance (DR) measurements, which allow us to provide a fair estimation of  $E_{\rm b}$ in our samples, also useful to estimate the changes in exciton radiative lifetime induced by graphene (see Sec.~\ref{SecTRPL}.1).

\subsection{BN capped MoSe$_2$/graphene}
\begin{figure*}[!h]
\begin{center}
\includegraphics[width=0.5\linewidth]{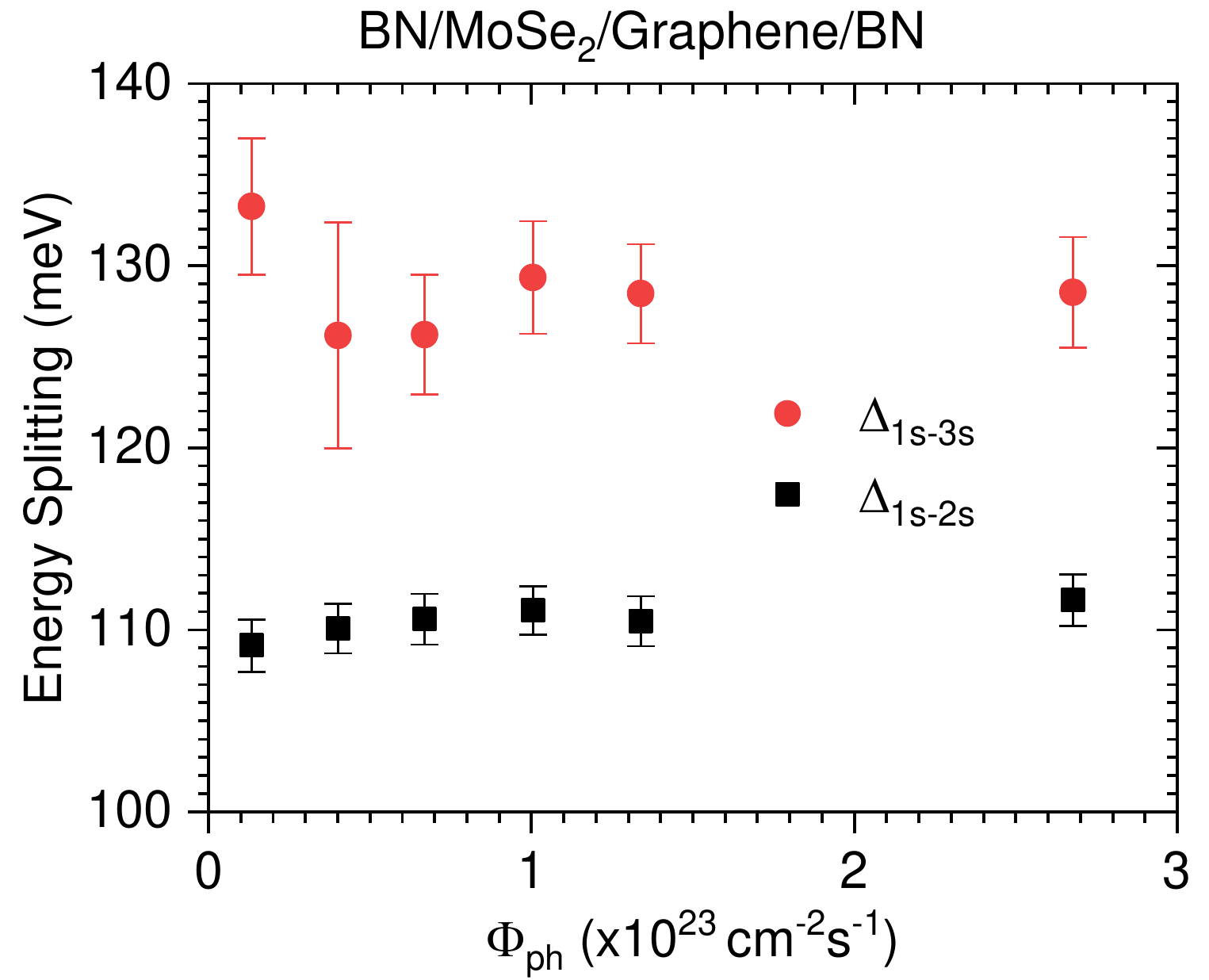}
\caption{\textbf{Dielectric screening and reduced exciton binding energy in BN-capped MoSe$_2$/graphene.} Splittings $\Delta_{1\rm s-2\rm s}$ (black squares) and $\Delta_{1\rm s -3\rm s}$ (red circles)  between the $\rm X^0_{1\rm s }$  and the $\rm X^0_{2\rm s}$ and $\rm X^0_{3\rm s}$ excitons, respectively. The data are extracted from the PL spectra shown in Fig.~3a and shown as a function of the \textit{cw} incoming photon flux at 2.33~eV and $\rm T=4~\rm K$.}
\label{FigS_1s2s3s}
\end{center}
\end{figure*}

First, on the BN-capped MoSe$_2$/graphene sample discussed in Fig.~2 and 3, we find splittings $\Delta_{1\rm s-2\rm s}=110 \pm 2~\rm {meV}$ between the $\rm X^0_{1\rm s}$ and $\rm X^0_{2\rm s}$ states and $\Delta_{1\rm s-3\rm s}=127 \pm 4~\rm {meV}$  between the $\rm X^0_{1\rm s}$ and $\rm X^0_{3\rm s}$ states (see fit in Fig.~3c and Fig.~\ref{FigS_1s2s3s}).

\subsection{BN-supported MoSe$_2$/graphene}

We also measured $\Delta_{1\rm s-2\rm s}\approx 113~\rm{meV}$ and $\Delta_{1\rm s -3\rm s}\approx 135~\rm{meV}$ using PL spectroscopy (as in Fig.~3 and~\ref{FigS_1s2s3s}) in the 1LG/MoSe$_2$/BN sample introduced in Sec.~\ref{SecGamma} (see Fig.~\ref{FigS_E2018_comp}). The $\rm X^0_{2\rm s}$ and $\rm X^0_{3\rm s}$ features nearly merge into one slightly asymmetric feature in 2LG/MoSe$_2$/BN and we estimate $\Delta_{1\rm s-2\rm s}\approx 109~\rm{meV}$. The reference region of the sample (vacuum/MoSe$_2$/BN) displays $\Delta_{1\rm s -2\rm s}\approx 153~\rm{meV}$ and its 3s feature merges with the low energy wing of the hot PL feature from the 1s B exciton ($\rm B_{1\rm s }$) that is near 210~meV above $\rm X^0$ in all parts of the sample. Interestingly, although this sample is not covered by a BN thin layer, these energy shifts are very similar to the values reported in Fig.~3 and~\ref{FigS_1s2s3s} in a BN-capped sample. These observations suggest that the screening experienced by a "bare" TMD monolayer deposited onto a transparent substrate experience is marginally affected by the presence of a top dielectric layer. Let us also stress that sample to sample variations in the energy splitting between excitonic states may reflect various levels of dielectric disorder, as recently investigated in details in Ref.~\onlinecite{Raja2019}.

\begin{figure*}[!h]
\begin{center}
\includegraphics[width=0.42\linewidth]{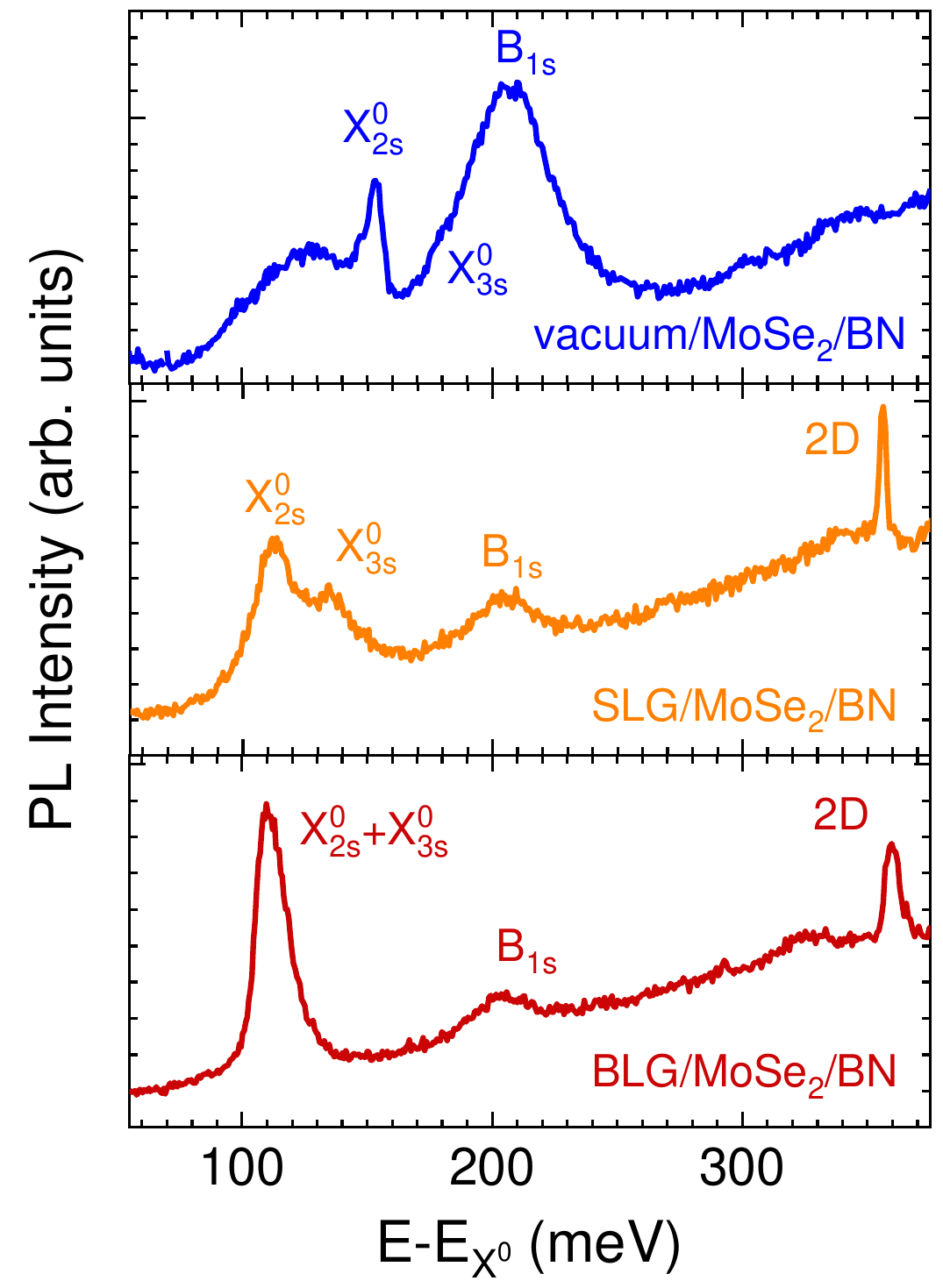}
\caption{\textbf{Dielectric screening and reduced exciton binding energy in BN/ MoSe$_2$/graphene.} Hot photoluminescence (similar to the data in Fig.~3) from the 2s ($\rm X^0_{2\rm s}$) and 3s ($\rm X^0_{3\rm s}$) A excitons as well as from the 1s B excitons in the BN-supported sample introduced in Sec.~\ref{SecGamma} and Fig.~\ref{FigS_E2018_comp}. Note that the $\rm X^0_{2\rm s}$ and $\rm X^0_{3\rm s}$ features merge in 2LG/MoSe$_2$/BN.  The energy scale is shown relative to the 1s exciton ($\rm X^0$) energy $E_{\rm X^0}$. The spectra were recorded at $\rm T=15~\rm K$ with a laser photon energy of 2.33~eV. The Raman 2D-mode features of single layer and bilayer graphene appear near  2.00 eV, i.e., around~370~meV above the $\rm X^0$ lines.}
\label{FigS_E2018_2s3s}
\end{center}
\end{figure*}

\subsection{BN-capped MoS$_2$/graphene}

\begin{figure*}[!h]
\begin{center}
\includegraphics[width=0.65\linewidth]{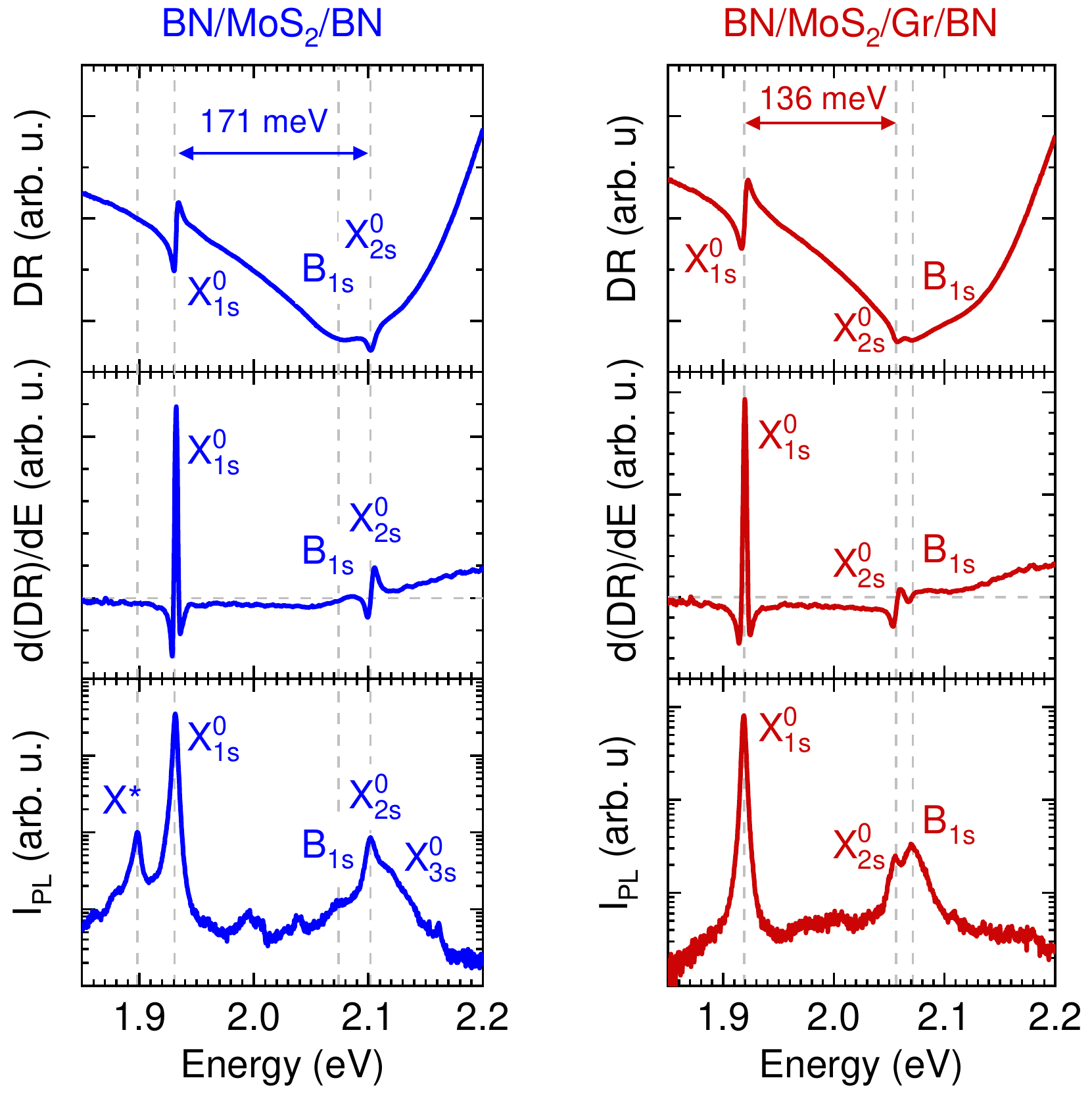}
\caption{\textbf{Dielectric screening and reduced exciton binding energy in BN-capped MoS$_2$/graphene.} From top to bottom: differential reflectance (DR), its derivative and photoluminescence (PL) spectra (shown in semilog scale) of a BN-capped MoS$_2$ sample (left, blue traces) and a neighboring BN-capped MoS$_2$/graphene region (right, dark red traces). The relevant excitonic states are indicated by vertical dashed lines. The horizontal dashed lines in the middle panels indicate the zero of the DR derivative. All measurements were done at $\rm T=15~\rm K$. The PL measurements were done with an incoming laser photon energy of 2.33~eV.}
\label{FigS_MoS2_spectro}
\end{center}
\end{figure*}

Similarly, using differential reflectance and PL spectroscopy, we could measure $\Delta_{1\rm s -2\rm s}\approx136~\rm{meV}$ and $\Delta_{1\rm s -2\rm s}\approx171~\rm{meV}$   in BN-capped MoS$_2$/graphene and BN-capped MoS$_2$ sample, respectively, as shown in Fig.~\ref{FigS_MoS2_spectro}.

\subsection{Discussion}
The measured $\Delta_{1\rm s -2\rm s}$ and $\Delta_{1\rm s -3\rm s}$ in TMD/graphene heterostructures must be compared with the larger splittings observed TMD samples (see Fig.~\ref{FigS_1s2s3s},~\ref{FigS_E2018_comp},~\ref{FigS_MoS2_spectro} and Ref.~\onlinecite{Raja2017,Han2018,Robert2018,Goryca2019}), confirming appreciable reduction of  $E_{\rm b}$  in TMD/graphene heterostructures.  To estimate $E_{\rm b}$ in TMD/graphene samples, we exploit a very recent study where $E_{\rm b}$  has been quantitatively mapped out in BN-capped TMD samples~\cite{Goryca2019}.  For MoSe$_2$ and MoS$_2$, $E_{\rm b}=231~\rm{meV}$ and $221~\rm{meV}$ were found, respectively, while the 1s-2s splittings were 168~meV and 170~meV, respectively. 

Assuming for simplicity that the exciton binding energy reduction is  proportional to the reduction of the $\Delta_{1\rm s -2\rm s}$, we can estimate binding energies $E_{\rm b}\approx 150~\rm{meV}$ in our BN-supported and BN capped MoSe$_2$/graphene samples and 176~meV  in our BN-capped MoS$_2$/graphene sample, respectively. Our results reveal that one single layer of graphene significantly reduces $E_{\rm b}$ by up to $\sim30~\%$ irrespective of the presence of a BN top layer. Our findings on $E_{\rm b}$ are summarised in Table~\ref{T2}.

\setlength{\tabcolsep}{0.5cm} 
\begin{table}[h!] 
\begin{center} 
\begin{tabular}{c c c c} 
\hline \hline\\[-3.5ex] 
Sample & $\Delta_{1\rm s -2\rm s}~\rm{(meV)}$ & $\Delta_{1\rm s -3\rm s}~ \rm{(meV)}$ & $E_{\rm b}~\rm{(meV)}$  \\[1.5ex] 
\hline 
BN/MoSe/BN (from Ref.~\onlinecite{Goryca2019}) & $168\pm 2$ & - & $231 \pm 3$   \\[0.5ex]
BN/Gr/MoSe$_2$/BN (Figs 2, 3 and \ref{FigS_1s2s3s}) &$110 \pm 2$&$127\pm 4$ & $\approx 150$   \\[0.5ex]
\hline
MoSe$_2$/BN (Fig.~\ref{FigS_E2018_2s3s})&$153 \pm2$ &$180 \pm 5$ & $\approx 210$  \\[0.5ex]
Gr/MoSe$_2$/BN (Fig.~\ref{FigS_E2018_2s3s}) &$113\pm 2$ &$135 \pm 5$ & $\approx 155$   \\[0.5ex]
\hline
BN/MoS$_2$/BN (Fig.~\ref{FigS_MoS2_spectro})& $171\pm2$ & $195\pm5$ & $\approx 221$ \\[0.5ex]
BN/Gr/MoS$_2$/BN  (Fig.~\ref{FigS_MoS2_spectro}) &$136\pm2$ & - & $\approx 176$ \\[0.5ex]

\hline \hline
\end{tabular}
\caption{\textbf{Reduced exciton binding energies in TMD/graphene heterostructures.} Energy splittings between $\rm X^0_{1\rm s }$ and $\rm X^0_{2\rm s}$ and (if measurable) $\rm X^0_{3\rm s}$ states in the samples studied in this work. The estimated exciton binding energies $E_{\rm b}$ are indicated.}
\label{T2}
\end{center}
\end{table}

\clearpage

\section{Estimating laser-induced heating and exciton densities}
\label{heating}

\begin{figure*}[!h]
\begin{center}
\includegraphics[width=0.72\linewidth]{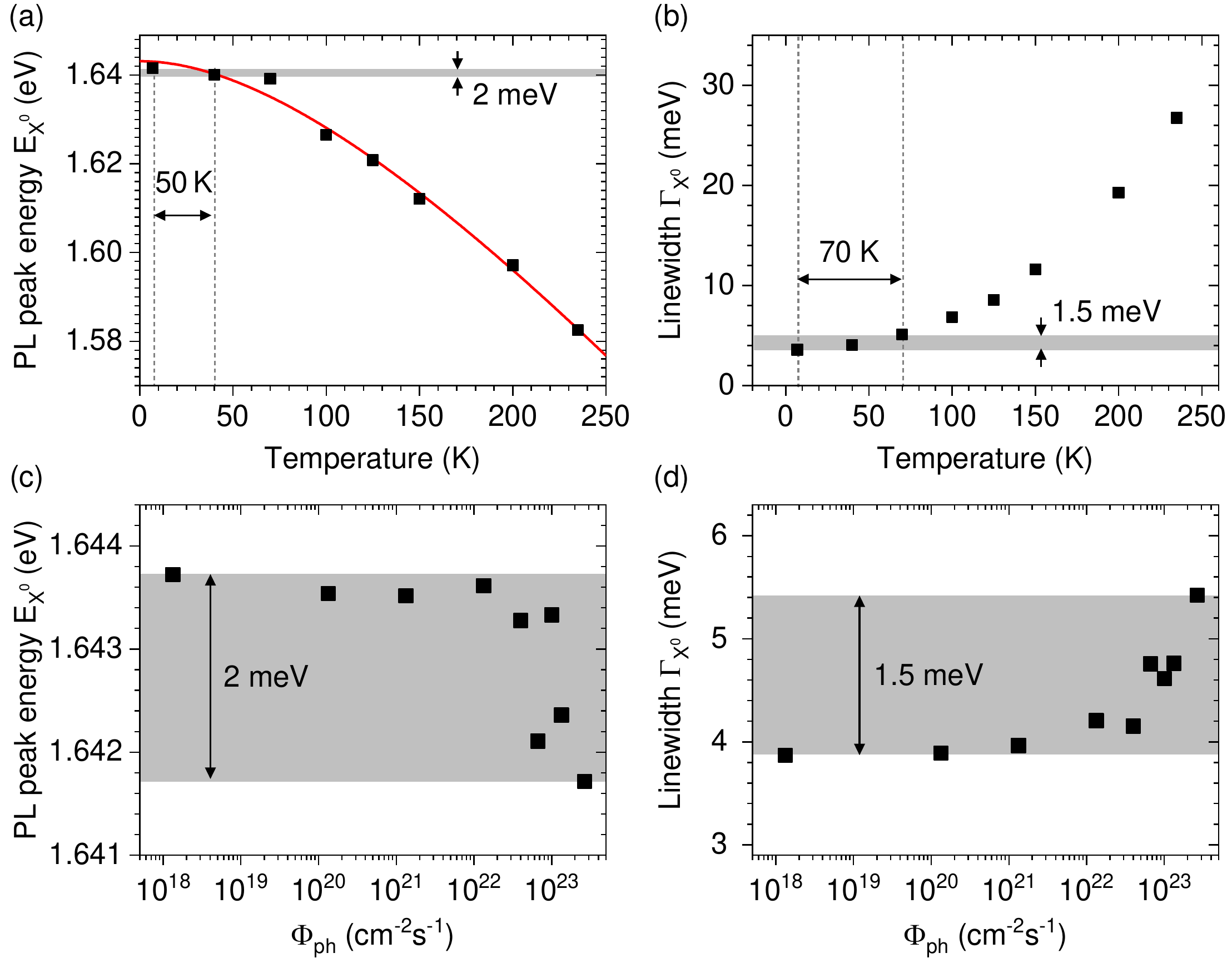}
\caption{\textbf{Estimating laser-induced heating.} Comparison between the $\rm X^0$ PL peak energy and linewidth (FWHM, $\Gamma_{\rm X}^0$) recorded as a function of the temperature set in our cryostat (a), and b), respectively) and as a function of the incident photon flux $\Phi_{\rm{ph}}$  for the BN-capped MoSe$_2$/graphene sample studied in Fig.~2-3 of the main manuscript. The slight excitonic redishift and broadening observed here allow us to  estimate that laser-induced sample heating is minimal and remains below 70~K at under the highest $\Phi_{\rm{ph}}$ employed here. The red line in (a) is a fit based on Varshni law.}
\label{FigS_heating}
\end{center}
\end{figure*}

In order to estimate the laser induced heating in our measurements, we have monitored the $\mathrm{X}^0$ energy and linewidth ($E_{\mathrm{X}^0}$  and $\Gamma_{\mathrm{X}^0}$) as a function of the incoming photon flux $\Phi_{\mathrm{ph}}$. As shown in Fig.~\ref{FigS_heating} (data extracted from the spectra in Fig.~3a), we only observe minimal $\mathrm X^0$ downshifts of at most 2~meV, associated with spectral broadening up to 1.5~meV. Considering the temperature dependence of $E_{\mathrm X^0}$  and $\Gamma_{\mathrm X^0}$, we can conservatively estimate that the temperature laser induced temperature rise at our sample is well below 30~K up to incoming photon flux $\Phi_{\mathrm{ph}}\sim 10^{23}~\mathrm{cm^2\ s^{-1}}$  and remains below  70~K at up to the highest values of $\Phi_{\mathrm{ph}}$ employed in our study.

Regarding the steady state excitonic density, we assume that we are in the linear recombination regime since the $\rm X^0$ PL intensity is linearly increasing with $\Phi_{\rm{ph}}$. Assuming an $\rm X^0$  lifetime of 2~ps (see Fig.~4 and Fig.~5) and an absorption coefficient near $10~\%$ at 2.33~eV~\cite{Li2014}, we can estimate maximal injected densities of hot excitons $n_{\rm{h}}\sim 2\times10^5 - 6\times 10^{10}~\rm{cm^{-2}}$) for the range of $\Phi_{\rm{ph}}$ employed in Fig.~3. 
Note also that time-resolved PL measurements were performed using picosecond oscillators with pulse fluence below a few $10^{12}~\rm{photons/cm^2}$. In these experiments, the laser photon energy was tuned between the 1s ($\rm X^0_{1\rm s}$) and 2s ($\rm X^0_{2\rm s}$) excitonic states. Assuming a conservative upper bound of $5~\%$ for the MoSe$_2$ monolayer absorptance in these conditions, we can estimate that $n_{\rm{h}}$ does not exceed $\sim 10^{11}~\rm{cm^{-2}}$, values for which we have verified that our samples were still in the linear regime.

\clearpage

\section{Supplementary discussion on exciton dynamics}
\label{SecTRPL}
In this section, we provide additional insights into the physical origin of the PL quenching discussed in the main text and give an estimate of the non-radiative exciton transfer time to graphene and of the $\rm X^0$ PL quantum yield.


As already introduced, the relaxation of $\rm X^0$ in TMD is essentially radiative with a short lifetime $\tau_{\rm X^{0}}^{\rm{rad}}$ of typically a few ps~\cite{Robert2016,Palummo2015,Fang2019}. In PL measurements, TMD samples cannot be resonantly excited at the $\rm X^0$ energy and the $\rm X^0$ state is populated after phonon-assisted relaxation of hot excitons (denoted $\rm X^{\rm h}$), on a timescale $\tau_{\rm {rel}}$  (see inset in Fig.~4a). Now, $\tau_{\rm {rel}}$ and hence the $\rm X^0$ formation and relaxation dynamics depend strongly on the environment. Due to a larger disorder, $\tau_{\rm {rel}}\sim 1~\rm {ps}$ or less in SiO$_2$ supported samples~\cite{Robert2016}, whereas $\rm X^{\rm h}$ relaxation can be as long as $\tau_{\rm {rel}}=18~\rm{ps}$ in state of the art BN-capped samples~\cite{Fang2019}. This value is significantly longer than $\tau_{\rm X^{0}}^{\rm{rad}}$, irrespective of modifications of $\tau_{\rm X^{0}}^{\rm{rad}}$ due to dielectric screening and to the Purcell effect~\cite{Fang2019}.  In addition, exciton dynamics and in particular $\rm X^{\rm h}$ relaxation \textit{a priori} depend on the incoming photon energy.

\subsection{Radiative liftetime and exciton transfer time}
\label{taurad}
For the sake of simplicity, in Fig.~4 of the main text, we have presented a low-temperature time-resolved PL (TRPL) study of a SiO$_2$ supported MoSe$_2$ monolayer partially covered by a graphene monolayer. In such a samples, the formation of $\rm X^0$ is faster than their radiative decay and cannot be experimentally resolved. Therefore, the TRPL measurements in Fig.~4 can safely be fit to the convolution of our IRF and an exponential decay.
We have found nearly equal \textit{excited state} lifetimes $\tau_{\rm X^0}\approx 2.3~\rm{ps}$ in graphene/MoSe$_2$/SiO$_2$ and in vacuum/MoSe$_2$/SiO$_2$ (see Fig.~4b,c), suggesting that the non-radiative transfer time of $\rm X^0$ excitons to graphene $\tau_{\rm G}^0$ is significantly longer than $\tau_{\rm X^0}^{\rm rad}$ in graphene/MoSe$_2$/SiO$_2$. Let us note, however, that due to dielectric screening, $\tau_{\rm X^0}^{\rm {rad}}$ is expected to be slightly longer in graphene/MoSe$_2$/SiO$_2$ than in vacuum/MoSe$_2$/SiO$_2$. As a result, observing similar $\tau_{\rm X^0}$ in graphene/MoSe$_2$/SiO$_2$ and in vacuum/MoSe$_2$/SiO$_2$ may be coincidental and result from a compensation between the increase of $\tau_{\rm X^0}^{\rm {rad}}$ and a contribution from non-radiative energy transfer of $\rm X^0$ with a timescale $\tau_{\rm{G}}^0\gtrsim 2.3 ~\rm{ps}$. 

As discussed in Sec.~\ref{Eb}, screening of the Coulomb interactions induced by semi-metallic graphene reduces the exciton binding energy or equivalently increases the exciton Bohr radius. Thus, the radiative lifetime, which scales as $E_{\rm b}^{-2}$, will increase due to the presence of graphene~\cite{Robert2016}. Taking the values shown in Table~\ref{T2}, we can estimate that $\tau_{\rm X^0}^{\rm {rad}}$ in BN-capped MoSe$_2$/graphene is approximately 2 times longer than in the bare MoSe$_2$ reference. A similar value is obtained for the BN/MoSe$_2$/graphene sample discussed in Fig.\ref{FigS_E2018_comp} and Fig.~\ref{FigS_E2018_2s3s} that is not covered by a top BN layer and hence has a layout similar to the SiO$_2$ supported sample discussed in Fig.~4. Considering the $\rm X^0$ radiative lifetimes $\tau_{\rm X^0}^{\rm {rad}}\approx 2.3-2.4~\rm{ps}$ in MoSe$_2$ monolayers (see Fig.~4 and Fig.~5), we estimate that $\tau_{\rm X^0}^{\rm {rad}}$ and the $\rm X^0$ exciton transfer time to graphene $\tau_{\rm G}^{0}$ are both close to 5~ps at cryogenic temperatures (here 15~K). Note that as described in the main text and in Sec.~\ref{SecTRPL}.3, the transfer of $\rm X^{\rm h}$   to graphene is likely to be more efficient than the transfer of zero momentum, optically active $\rm X^0_{1\rm s }$.

Let us point out that the presence of a single layer of graphene with larger refractive index than BN may further increase $\tau_{\rm X^0}^{\rm {rad}}$. However, as discussed in Ref.~\cite{Froehlicher2018,Fang2019}, the inclusion of an atomically thin layer in a multi-layered system has a negligible effect on the local field enhancement experienced by the TMD monolayer and we can safely consider that changes in $\tau_{\rm X^0}^{\rm {rad}}$ in TMD/graphene heterostructures essentially stem from screening of the Coulomb interactions induced by graphene.

\subsection{Emission quantum yield of the $\rm X^{0}$ state}
\label{Seceta}
We first define the total emission quantum yield $\eta_{\rm{tot}}$  as the number of emitted photons divided by the number of absorbed photons. The latter is defined as: $\eta_{\rm{tot}}=\eta_{\rm X^0}\eta_{\rm F}$, with $\eta_{\rm X^0}$, the number of emitted photons divided by the number of cold $\rm X^0$ (i.e, $\rm X^{0}_{1\rm s }$ excitons with center of mass momentum within the light cone)  and $\eta_{\rm F}$ the formation yield of cold $\rm X^0$  (number of cold $\rm X^0$ / number of absorbed photons). In practice, due to the limited collection efficiency in cryogenic optical setups and uncertainty in the sample absorptance, it is extremely challenging to measure $\eta_{\rm{tot}}$ accurately at low temperatures. We can, however, estimate $\eta_{\rm X^0}$.

Indeed, it was established in Ref.~\onlinecite{Robert2016,Fang2019} that at cryogenic temperatures (typically below 90 K), $\rm X^0$  recombination in MoSe$_2$ monolayers is essentially radiative such that the measured $\rm X^0$ lifetime $\tau_{\rm X^0}$ can be attributed to the radiative lifetime $\tau_{X^0}^{\rm{rad}}$. In other words $\eta_{\rm X^0}$ is very near $100 \%$ at low temperatures in MoSe$_2$ monolayers. Now, using the results in Fig.~4 and Fig.~5 and the discussion in Sec.~\ref{SecTRPL}.1, one obtains $\eta_{\rm X^0}^{~}=\tau_{\rm G}^{0}/\left(\tau_{\rm G}^{0}+\tau_{\rm X^0}^{\rm {rad}}\right)\sim 50\%$ both in SiO$_2$ supported and BN-capped MoSe$_2$/graphene heterostructures.
Besides, the yield $\eta_{\rm{tot}}$ also depends strongly on whether the TMD under study is a bright or dark material. In the case of bright (e.g., MoSe$_2$) TMDs, $\eta_{\rm F}$ is limited by exciton transfer to graphene (in the case of graphene/TMD heterostructures) and by the presence of residual dopants (in the case of bare TMD), traps and neutral defects (in both TMD and TMD/graphene). In the case of dark (here W-based) TMDs, the $\rm X^0$ population is also limited by the presence of lower lying dark states ($\rm X^{\rm D}$) that can be efficiently populated from the $\rm X^0$ state. In WS$_2$/graphene and WSe$_2$/graphene heterostructures, formation of $\rm X^{\rm D}$ is also limited by $\rm X^{\rm h}$ transfer to graphene. As a result, $\eta_{\rm X^0}$ and $\eta_{\rm {tot}}$ will be intrinsically smaller in W-based TMDs than in MoSe$_2$, whether or not these TMDs are coupled to graphene.

\subsection{Photoluminescence excitation spectroscopy}
\label{SecPLE}

In Fig.~4 and Fig.~5, care has been taken to investigate exciton dynamics using excitation only 100-110~meV above the $\rm X^0$ line in MoSe$_2$, i.e., below or close to the $\rm X^{0}_{2\rm s}$ line. This configuration is the same as in Ref.~\onlinecite{Robert2016} and~\onlinecite{Fang2019}. Noteworthy, it was reported in Ref.~\onlinecite{Fang2019} that exciton dynamics in BN-capped MoSe$_2$ and in particular the hot exciton relaxation time (i.e., the $\rm X^0$ formation time) is independent on the energy difference $\Delta$ between the $\rm X^0$ line and the incoming photon energy for $\Delta=7-110 ~\rm{meV}$.

\begin{figure*}[!ht]
\begin{center}
\includegraphics[width=0.77\linewidth]{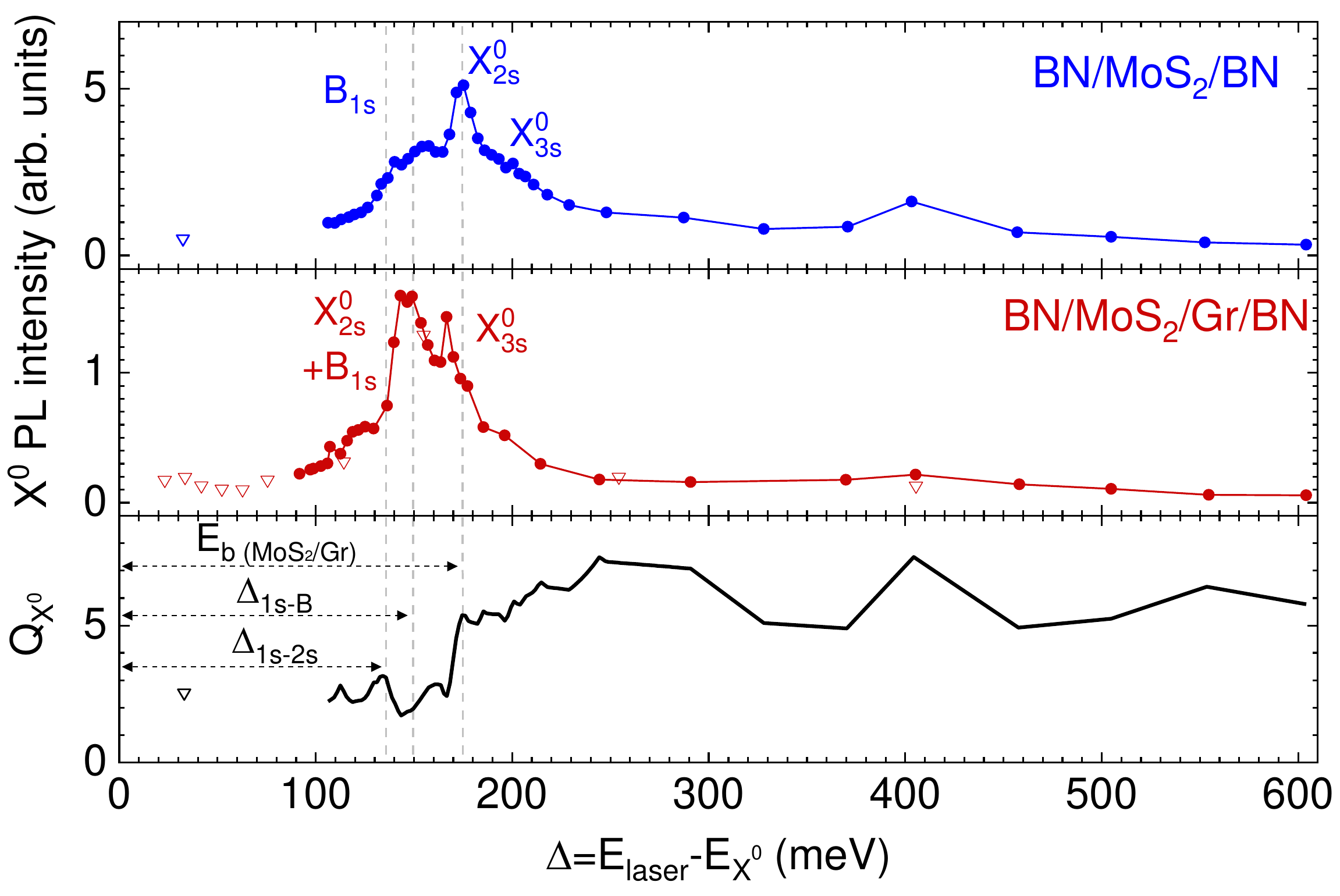}
\caption{\textbf{Photoluminescence excitation spectroscopy (PLE) of the sample introduced in Fig.~\ref{FigS_MoS2_spectro}.} Integrated intensity of the $\mathrm X^0$ PL line as a function of $\Delta$, the energy shift between the incoming photon energy and $\mathrm X^0$ PL line (near 1.93~eV) in BN-capped MoS$_2$ (top panel), and BN-capped MoS$_2$/graphene (middle panel).  Open symbols correspond to a second measurement run on the same sample. The bottom panel shows the $\mathrm X^0$ PL quenching factor $Q_{\mathrm X^0}$ (see also Table~\ref{T1}). Since the $\mathrm X^0$ lines have slightly distinct energies in MoS$_2$ and MoS$_2$/graphene, $Q_{\mathrm X^0}$ is evaluated after interpolation of the PLE data in the top panel. The energy shifts between $\mathrm X^0_{1\mathrm s }$ and the $\mathrm X^0_{2\mathrm s}$ and $\mathrm B_{1\mathrm s }$ as well as $E_{\mathrm b}$ are indicated in the bottom panel for MoS$_2$/graphene.  All data were recorded at a temperature of $\mathrm{T}= 15~\mathrm{K}$.}
\label{FigS_MoS2_PLE}
\end{center}
\end{figure*}

To get more insights into $\rm X^{\rm h}$ dynamics and PL quenching, we have performed photoluminescence excitation spectroscopy (PLE) and extracted the $\rm X^0$ PL quenching factor $Q_{\rm X^0}$ (see Table~\ref{T1}) as a function of $\Delta$. We have chosen to focus our study on BN-capped MoS$_2$/graphene sample shown in Fig.~\ref{FigS_MoS2_spectro}. In contrast to bare MoSe$_2$, which shows sizeable emission from $\rm X^{\star}$ (see~Fig.~4 and~Fig.~5), our bare MoS$_2$ references display minimal $\rm X^{\star}$ emission (see Fig.~1 and Fig.~\ref{FigS_MoS2_spectro}), which allows us to compare $\rm X^0$ dynamics in MoS$_2$ and MoS$_2$/graphene without considering a significant contribution from $\rm X^{\star}$. 
Our main results are shown in Fig.~\ref{FigS_MoS2_PLE}. PLE spectroscopy permits to identify the $\rm X^0_{2\rm s}$ and $\rm B_{1\rm s }$ excitons, that have unambiguously been assigned by means of DR and PL measurements (see Fig.~\ref{FigS_MoS2_spectro}). The quenching factor $Q_{\rm X^0}$  reaches low values $\sim 2$ for $\Delta\approx 100~\rm{meV}$, i.e., significantly below $\rm X^0_{2\rm s}$ and remains roughly constant down to $\Delta\approx 33~\rm {meV}$. Except for strictly resonant excitation of excited excitonic states (e.g., $\rm X_{2\rm s}^0$), no optically active excitons can be formed through vertical transitions. Instead, hot 1s excitons with finite center of mass momentum are formed through acoustic and optical phonon assisted processes and relax to form cold $\rm X^0$ within the light cone. We attribute the residual PL quenching to non-radiative transfer of hot and cold excitons to graphene. Remarkably, $Q_{\rm X^0}$ rises sharply up to approximately 5 to 7 above $\Delta \approx 175 ~\rm{meV}$, a value that precisely coincides with the $\rm X_{2\rm s}^0$ state in bare MoS$_2$ but also with the exciton binding energy in MoS$_2$/Graphene (as shown in Table~\ref{T2}). Above this threshold, $Q_{\rm X^0}$ does not significantly increase further as $\Delta$ augments up to $600~\rm{meV}$. These results suggest that the rise of $Q_{\rm X^0}$  near $\Delta \approx 175 ~\rm{meV}$ cannot solely be assigned to a resonance with the $\rm X_{2\rm s}^0$ state in MoS$_2$ but suggests that $\rm X^{\rm h}$ generated above the free carrier continuum associated with the A excitonic manifold (for instance, hot B excitons) have a higher transfer efficiency to graphene than hot $\rm X^0_{1\rm s }$ excitons.

\clearpage

\section{PL of a TMD monolayer onto a metal thin film.}
\label{Metal}
Using a 2D metallic layer such as graphene instead of a thin film of bulk metal is certainly advantageous in view of applications. Our work establishes that a single atomic layer of graphene warrants highly efficient filtering based on selective energy transfer. Nevertheless, thin films of bulk metals, such as Au, Ti, Pt\dots are routinely used as electrical contacts to 2D materials in electronic and optoelectronic devices and deserve special attention. Metallic thin films will be unavoidably rougher than 2D materials all the more so if the 2D materials are deposited on Boron Nitride (BN). As a result, van der Waals coupling between a metal film and an atomically thin semiconductor will  be weaker than within a van der Waals heterostructure and the metal/TMD interface will be much more inhomogeneous than a van der Waals heterointerface. Such inhomogeneities are prone to affect the emission characteristics of the TMD monolayer.

\begin{figure*}[!h]
\begin{center}
\includegraphics[width=0.85\linewidth]{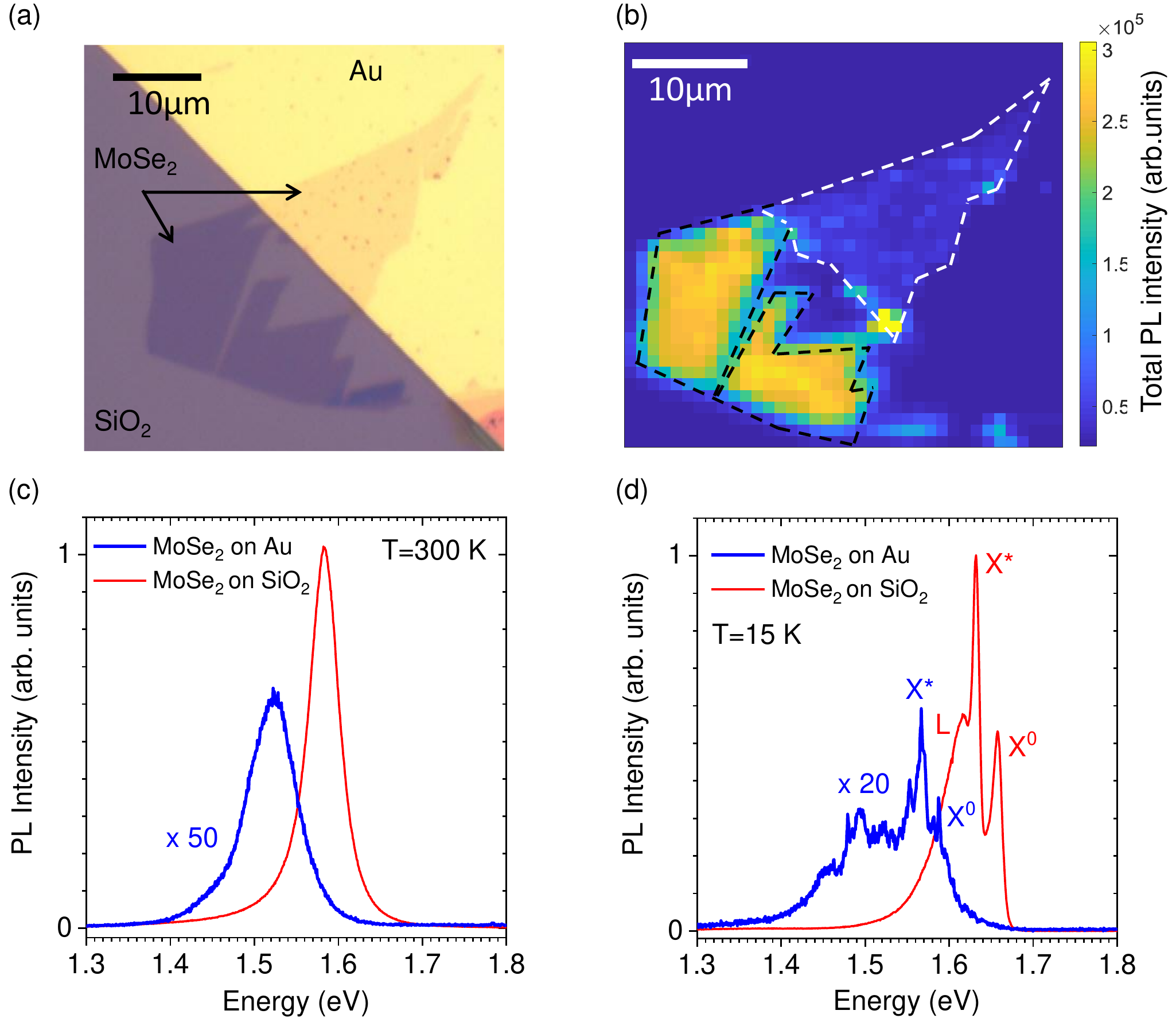}
\caption{\textbf{Photoluminescence spectroscopy of $\mathrm{MoSe_2}$ on a gold substrate.} (a) Optical and (b) room temperature photoluminescence (PL) images of a bare $\mathrm{MoSe_2}$ monolayer deposited at the border between a $\mathrm{SiO_2/Si}$ substrate and a gold (Au) thin film. (c) and (d) PL spectra of  $\mathrm{MoSe_2}$ on $\mathrm{SiO_2}$ (red) and on Au (blue) recorded at room temperature (c) and at 15 K (d). ``L'' indicates a sizeable contribution from localised states giving rise to a low-energy shoulder in the reference spectrum on $\mathrm{SiO_2}$ that gives a prominent feature in the PL spectrum of  $\mathrm{MoSe_2}$ on gold.}
\label{FigS_Au}
\end{center}
\end{figure*}

As shown in Fig.~\ref{FigS_Au}, we have performed PL experiments on a MoSe$_2$ Monolayer deposited onto a 60~nm-thick Au film and compared the PL spectrum of MoSe$_2$ on Au to that of a neighbouring region of the  MoSe$_2$ flake lying onto SiO$_2$. At room temperature, we observe large PL quenching (akin to previous findings~\cite{Froehlicher2018,He2014}) and significant broadening of the PL spectra suggesting nanoscale variations of the coupling between MoSe$_2$ and Au. At low temperature, we observe a complex, even broader PL spectrum, where the $\rm X^0$ and $\rm X^{\star}$ lines can tentatively be identified. In stark contrast with the case of TMD/graphene heterostructures, we still observe strong $\rm X^0$ PL quenching, sizeable $\rm X^{\star}$ emission, as well as evidence for disorder, inhomogeneous broadening and exciton localization, as shown in Fig.~\ref{FigS_Au}d. In other words, while the PL spectrum of MoSe$_2$/graphene only reveals intrinsic $\rm X^0$ emission, even when the MoSe$_2$ layer lies on SiO$_2$ (see Fig.~4 and Table \ref{T1}), MoSe$_2$ on Au displays prominent extrinsic PL features. It is therefore clear that using layered metals, ideally in the 2D limit is undoubtedly preferable than using bulk metals in view of the photonic and optoelectronic applications based on the concepts introduced in our manuscript.

\clearpage

\end{document}